\documentclass[
  journal=pasa,
  manuscript=research-paper, 
  year=202X,
  volume=YY,
]{cup-journal}

\usepackage{amsmath}
\usepackage{amssymb,microtype,siunitx,booktabs}
\usepackage{pdflscape}
\usepackage{hyperref}
\hypersetup{colorlinks,citecolor=blue,linkcolor=blue,urlcolor=blue}

\usepackage[nameinlink,capitalise]{cleveref}
\crefname{section}{Sect.}{Sects.}
\Crefname{section}{Section}{Sections}
\crefname{figure}{Fig.}{Figs.}
\Crefname{figure}{Figure}{Figures}
\crefname{equation}{Eq.}{Eqs.}
\Crefname{equation}{Equation}{Equations}
\crefname{table}{Table}{Tables}
\crefname{appendix}{Appendix}{Appendices}

\newcommand{\hi}{\mbox{H\,{\sc i}} } 

\title{The ASKAP-FLASH survey: A first look at the multiwavelength properties and redshift distribution of compact radio sources}

\author{W. Roster}
\affiliation{Max-Planck-Institut f\"ur extraterrestrische Physik, Giessenbachstr. 1, 85748 Garching, Germany}
\email[W. Roster]{wroster@mpe.mpg.de}

\author{E. M. Sadler}
\affiliation{Sydney Institute for Astronomy, School of Physics A28, University of Sydney, NSW 2006, Australia}
\alsoaffiliation{ARC Centre of Excellence for All-Sky Astrophysics in 3 Dimensions (ASTRO 3D)}
\alsoaffiliation{ATNF, CSIRO, Space and Astronomy, PO Box 76, Epping, NSW 1710, Australia}

\author{E. K. Mahony}
\affiliation{ATNF, CSIRO, Space and Astronomy, PO Box 76, Epping, NSW 1710, Australia}

\author{M. Salvato}
\affiliation{Max-Planck-Institut f\"ur extraterrestrische Physik, Giessenbachstr. 1, 85748 Garching, Germany}

\author{H. Yoon}
\affiliation{Institute for Data Innovation in Science, Seoul National University, 1 Gwanak-ro, Gwanak-gu, Seoul 08826, Republic of Korea}
\alsoaffiliation{Astronomy Program, Department of Physics and Astronomy, Seoul National University, 1 Gwanak-ro, Gwanak-gu, Seoul 08826, Republic of Korea}
\alsoaffiliation{Korea Astronomy and Space Science Institute, 776 Daedeokdae-ro, Daejeon 34055, Republic of Korea}
\alsoaffiliation{Sydney Institute for Astronomy, School of Physics A28, University of Sydney, NSW 2006, Australia}

\author{M. Kluge}
\affiliation{Max-Planck-Institut f\"ur extraterrestrische Physik, Giessenbachstr. 1, 85748 Garching, Germany}

\author{R. Shirley}
\affiliation{Max-Planck-Institut f\"ur extraterrestrische Physik, Giessenbachstr. 1, 85748 Garching, Germany}

\author{E. Kerrison}
\affiliation{Sydney Institute for Astronomy, School of Physics A28, University of Sydney, NSW 2006, Australia}
\alsoaffiliation{ARC Centre of Excellence for All-Sky Astrophysics in 3 Dimensions (ASTRO 3D)}
\alsoaffiliation{ATNF, CSIRO, Space and Astronomy, PO Box 76, Epping, NSW 1710, Australia}

\author{J. Buchner}
\affiliation{Max-Planck-Institut f\"ur extraterrestrische Physik, Giessenbachstr. 1, 85748 Garching, Germany}

\author{Z. Igo}
\affiliation{Max-Planck-Institut f\"ur extraterrestrische Physik, Giessenbachstr. 1, 85748 Garching, Germany}

\author{R. Davies}
\affiliation{Centre for Astrophysics and Supercomputing, Swinburne University of Technology, Hawthorn, Victoria 3122, Australia}
\alsoaffiliation{ARC Centre of Excellence for All-Sky Astrophysics in 3 Dimensions (ASTRO 3D)}

\author{J. R. Allison} 
\affiliation{First Light Fusion Ltd., Unit 9/10 Oxford Pioneer Park, Mead Road, Yarnton, Kidlington OX5 1QU, UK}

\author{S. S. Shabala}
\affiliation{School of Natural Sciences, Private Bag 37, University of Tasmania, Hobart 7001, Australia}
\alsoaffiliation{ARC Centre of Excellence for All-Sky Astrophysics in 3 Dimensions (ASTRO 3D)}

\author{V. A. Moss}
\affiliation{ATNF, CSIRO, Space and Astronomy, PO Box 76, Epping, NSW 1710, Australia}
\alsoaffiliation{Sydney Institute for Astronomy, School of Physics A28, University of Sydney, NSW 2006, Australia}

\author{H. Starck}
\affiliation{Max-Planck-Institut f\"ur extraterrestrische Physik, Giessenbachstr. 1, 85748 Garching, Germany}

\author{M. Whiting}
\affiliation{ATNF, CSIRO, Space and Astronomy, PO Box 76, Epping, NSW 1710, Australia}

\author{K. Nandra}
\affiliation{Max-Planck-Institut f\"ur extraterrestrische Physik, Giessenbachstr. 1, 85748 Garching, Germany}

\author{J. Weller}
\affiliation{Max-Planck-Institut f\"ur extraterrestrische Physik, Giessenbachstr. 1, 85748 Garching, Germany}

\affiliation{Max-Planck-Institut f\"ur extraterrestrische Physik, Giessenbachstr. 1, 85748 Garching, Germany}
\alsoaffiliation{LMU Munich, Universit\"ats-Sternwarte, Scheinerstr. 1, 81679 M\"unchen, Germany}

\received {dd Mmm YYYY}
\revised  {dd Mmm YYYY}
\accepted {dd Mmm YYYY}
\published{22 September 202X}


\begin{document}

\begin{abstract}
We present the characterisation, including a photometric redshift (photo-$z$) analysis, of the optical counterparts (CTPs) to over 45\,000 bright ($S_{856\,\rm MHz} \geq$ 30\,mJy) compact radio sources, identified across all ASKAP First Large Absorption Survey in \hi (FLASH) fields observed up to April 2025. These sources constitute a large, homogeneous population of background continuum sightlines specifically selected to enable statistical studies of cold gas at intermediate redshifts of $0.42 \leq z \leq 1$. As spectroscopic redshift measurements are not available for the majority of these candidate absorbers, we estimate photo-$z$s for the CTPs of all FLASH continuum sources cross-matched to the tenth data release of the DESI Legacy Imaging Surveys (LS10). Using these estimates, we establish the redshift distribution and find that approximately 13\% of continuum sources lie at $z<0.42$ (foreground), 35\% within the detectability range of FLASH (`in-band'), and 52\% at $z>1$ (background). We examine the subset of FLASH continuum sources with CTPs in the eROSITA X-ray survey, providing additional insight into their AGN content, multiwavelength properties, and environments. Finally, we discuss how this information can be used as a statistical prior to aid in distinguishing between associated and intervening \hi absorption systems and estimating the total comoving absorption path length of the survey, establishing a framework for incorporating redshift-based priors in future large radio absorption surveys. We release a catalogue of LS10 counterparts to FLASH continuum sources, providing photo-$z$ estimates, associated uncertainties, and measures of redshift degeneracies.
\end{abstract}

\section{Introduction}
\label{sec:int}

As in most domains of extragalactic astronomy, redshift ($z$) plays a critical role in radio surveys by linking measured quantities of galaxies to their underlying physical properties and evolutionary histories. However, determining the redshift distribution $n(z)$ of a galaxy sample selected purely on the basis of radio flux density remains a persistent challenge. In addition to flux variability, the key difficulty arises from the fact that the apparent brightness of a radio source is largely uncorrelated with its distance \citep[][]{hoyle1966,bolton1966,norris2019}. This is due to the extremely broad radio luminosity function (RLF) of galaxies, which spans more than six orders of magnitude \citep{mauch2007}. As a result, some of the brightest radio sources in the sky are, in fact, located at high redshifts ($z > 1$), powered by extremely luminous radio-loud (RL) active galactic nuclei (AGN). Blind radio continuum surveys, unlike many optical or infrared (IR) imaging surveys, almost never provide redshift information on their own, as reliable estimation in this regime is particularly challenging since dominant synchrotron emission produces a smooth, featureless spectrum without strong spectral markers. In most cases, redshift estimation requires the identification of a multi-wavelength counterpart (CTP) and, ideally, the acquisition of an (optical) emission line spectrum. This process can be both observationally expensive and technically difficult, particularly for the high-$z$ RL AGN population, which, partially due to dust extinction, is often faint at optical and IR wavelengths \citep[][]{debreuck2002,norris2019}. In contrast, radio selection is particularly suitable for detecting AGN with weak nuclear activity, helping to build a more complete AGN census \citep[][]{Delvecchio_2017, Radcliffe_2021, Mazzo_2026}. 

\begin{figure*}[t!]
\centering
\includegraphics[angle=0,width=0.9\hsize]{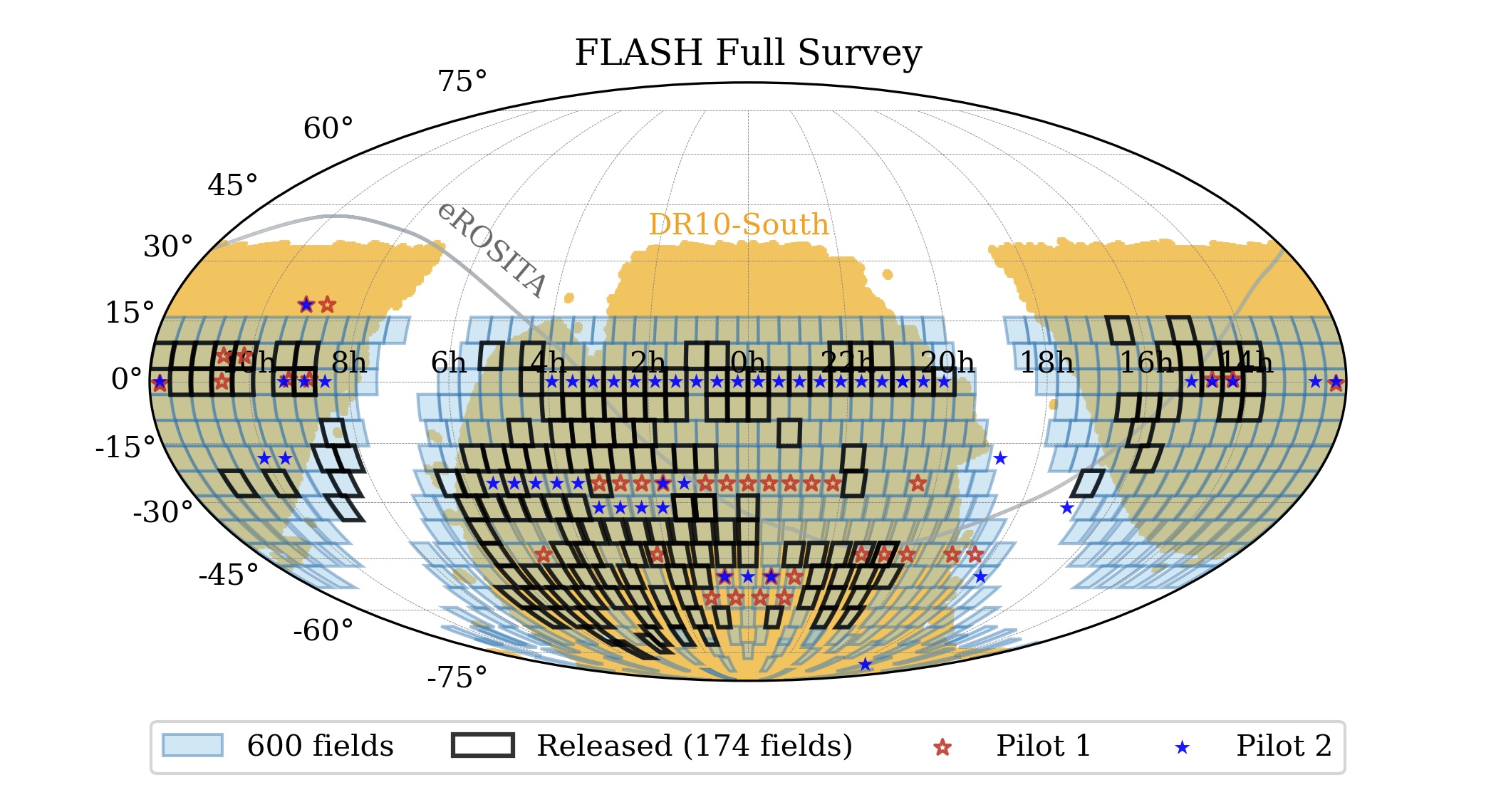}
\caption{FLASH survey footprint in equatorial coordinates. Black boxes indicate the fields observed up to April 2025, with those included in the pilot surveys marked by star symbols. The eROSITA-DE reference of the western Galactic hemisphere and the coverage of tenth release of the Dark Energy Spectroscopic Instrument (DESI) Legacy Imaging Surveys South (DR10-South) are overplotted for context.}
\label{fig:intro}
\end{figure*}

The challenge of determining the $n(z)$ of a flux-limited sample of radio sources can be tackled using either a modelling or an empirical approach. The modelling route \citep{dunlop1990, dezotti2010} relies on combining observed radio source counts with RLFs for various source classes, alongside assumptions about their redshift evolution. This framework yields a statistical prediction of the characteristic $n(z)$ and enables the creation of `mock' catalogues for survey planning and interpretation \citep[e.g.][]{wilman2008, bonaldi2019}. A major strength of this approach lies in its flexibility, offering a broad population-level perspective that can be updated as new data become available \citep[e.g.][]{lin2024}. However, it does not provide redshift estimates or physical details for individual sources. In contrast, the empirical method \citep{laing1983, McCarthy1996, best1999} involves directly associating each radio detection with an optical CTP and obtaining a spectroscopic redshift (spec-$z$). This enables the study of individual radio sources and their host galaxies in detail. Nevertheless, the approach is resource-intensive, particularly for faint, high-$z$ sources, such that comprehensive spectroscopic coverage is typically restricted to a relatively small subset of (bright) radio sources. Until ongoing and upcoming wide-field, multi-object spectroscopic facilities, such as the Prime Focus Spectrograph \citep[PFS,][]{Tamura_2016}, the 4-metre Multi-Object Spectroscopic Telescope \citep[4MOST,][]{Dejong_2019,Duncan_2023}, and the Sloan Digital Sky Survey \citep[SDSS-V,][]{Juna_2026} deliver large, statistically complete samples with secure optical redshifts, progress toward characterising the full radio population must rely on alternative methods. In this context, photometric redshift (photo-$z$) techniques \citep[e.g.][]{norris2019, manning_2020, Zhou_2021, Newman_2022, hatfield_2022, duncan2022, Luken_2023, Merz_2025}, provide a practical and scalable means of extending redshift information to $z \sim 1-2$ across larger and fainter radio samples, particularly for AGN with multi-band optical or IR CTPs \citep[see][for a recent application]{hardcastle2025}. Recently, \cite{roster2024} developed a new approach to photo-$z$ estimation called {\sc{PICZL}}, which incorporates spatial information in addition to traditional colour features. By operating directly on imaging data, the method leverages morphological properties and compactness-related radial surface-brightness profiles to help break colour–redshift degeneracies and improve overall photo-$z$ accuracy. The paper demonstrated that, when applied to AGN in wide-area optical surveys (>1000 deg\textsuperscript{2}) and using comparable photometric inputs, {\sc PICZL} outperforms both similar machine-learning (ML) and template-fitting approaches. In this paper, we employ {\sc{PICZL}}, to determine the expected $n(z)$ of a sample of compact\footnote{Sources smaller than about 10-15\,arcsec in size, which make up at least 90\% of radio AGN at $z>0.4$ \citep[][]{Ching2017, Baldi23}} radio sources with flux densities $S_{856\, \rm{MHz}} >$ 30\,mJy. The motivation for characterising this population of relatively bright radio sources arises from its role in the Australian square kilometre array pathfinder \citep[ASKAP,][]{Hotan_2021} First Large Absorption Survey in \hi \citep[FLASH;][]{allison22,Yoon_2025}, which uses these sources as probes to search for redshifted \hi 21-cm absorption lines over the range $0.42 < z < 1.0$. Depending on their redshift, sources may fall into three distinct categories: i) foreground sources at $z < 0.42$, where the redshift is too low for FLASH to detect \hi absorption; ii) sources in the redshift range $0.42 < z < 1.0$, where FLASH can potentially detect either associated \hi absorption features or intervening \hi along the line of sight; and iii) background sources at $z > 1$, where intervening \hi absorption can be detected across the full FLASH redshift range. Accurately characterising the $n(z)$ distribution is therefore essential to estimate the redshift pathlength probed by FLASH for different kinds of \hi absorption lines and interpreting the distribution of absorbers in a statistical sense \citep[][]{Allison_21}. Beyond enabling reliable distance estimates, redshift information is essential to unlocking the full scientific potential of FLASH, including the identification and characterisation of AGN and star-forming galaxies (SFGs), the interpretation of radio power and radio loudness distributions, and the study of obscuration and host-galaxy properties. Combined with multiwavelength data, one can additionally explore the X-ray and infrared properties of radio-selected sources, as well as potential ties in regard to their environments. In this work, we take a first step toward demonstrating the types of population-level analyses that become possible with FLASH once appropriate redshift information is available.

The paper is organised as follows: \Cref{sec:flash_z} describes the FLASH radio survey and the respective source sample. In \Cref{sec:piczl}, we introduce {\sc PICZL}, an image-based photo-$z$ estimation code. \Cref{sec:tests} presents an evaluation of {\sc PICZL}’s performance on radio-selected AGN. Our selection function
and identification of optical CTPs are presented in \Cref{sec:CSS} and \ref{app:1}, respectively. We show applications through our multiwavelength analyses in \Cref{sec:red_results} and assess whether and how photo-$z$s can be used to improve the classification of \hi absorption systems in \Cref{sec:improved_selection}. An outlook in regard of our findings is presented in \Cref{sec:out} with a concluding summary in \Cref{sec:summary}. We express magnitudes in the AB system \citep{oke_1983} and assume cosmological parameters of $H_0=70\,$km\,s$^{-1}$\,Mpc$^{-1}$, $\Omega_{\rm M}=0.3$, $\Omega_{\Lambda}=0.7$ and $\Omega_{\rm{k}} = 0$.

\section{The ASKAP-FLASH radio source sample}
\label{sec:flash_z}

FLASH uses the wide-field ASKAP radio telescope \citep{hotan21} to search for redshifted \hi absorption lines tracing cool, neutral gas in the frequency range $712-1000$\,MHz, which corresponds to redshift $0.42<z<1$. To optimise the number of absorption lines detected, the survey is designed to be relatively shallow (two hours per pointing) and cover 24\,000$\deg^2$ of sky \citep[for survey details see][]{allison22}. A single FLASH observation covers $\sim 40$\,deg$^2$ of sky at uniform sensitivity, with a typical root-mean-square (rms) noise of 90\,$\mu$Jy/beam in the continuum images and 5.5\,mJy/beam per 18.5\,kHz spectral channel in the spectral-line cubes \citep{Yoon_2025}.

\subsection{The FLASH continuum source catalogues}
As the survey is still in progress, this study utilises only the continuum source catalogues for the first 174 of 600 individual fields observed by FLASH up to April 2025 (see the black boxes in \Cref{fig:intro}). The data products from FLASH are available at CSIRO ASKAP Science Data Archive (CASDA) public archive\footnote{\url{https://research.csiro.au/casda/}} and are described in detail by \cite{Yoon_2025}. For each FLASH field, a component catalogue and a separate island catalogue are produced by the Selavy source finder \citep{whiting2012}. 
It is important to understand the difference between these two catalogues. Each entry in the component catalogue corresponds to a 2D-Gaussian fit to an individual radio source identified by Selavy. In contrast, the island catalogue associates one or more components that are joined at some flux level and combines them into a single composite source whose total flux density and fitted radio centroid are catalogued. Typically, this means each entry in the island catalogue corresponds to one, physically distinct source on the sky, while the corresponding entries in the component catalogue capture its morphological structure. However, it should be mentioned that an island may contain more than one physical source, or a physical source may consist of more than one island.

To search for \hi absorption lines, a spectrum is extracted at the position of each catalogued component with a peak flux density $S_{856} \geq 30$\,mJy. A Bayesian line finder \citep{allison12} is then used to search for lines, estimate the statistical significance of each line, and fit parameters for the redshift, optical depth and velocity width of each line as described by \cite{Yoon_2025}. For a single, isolated radio source that is small compared to the 10-15\,arcsec ASKAP beam (which we term a `compact' source for the purpose of this paper), the component catalogue will contain a single component with the island and component positions agreeing well. Consequently, for a true point source, the peak flux density is approximately equal to the integrated flux density. In this case, for a compact radio source where we assume the emission to emerge from close to its core/galactic center, the position where \hi absorption can be detected is very likely the same as the position of the expected optical CTP of the radio continuum source. 

For a multi-component island, the situation is more complex. These sources are usually extended double or triple radio sources where the individual components are separated on the sky and the optical counterpart may not lie at the position of any of the individual radio components.  It is possible to detect intervening \hi absorption against compact lobes, hotspots or jet knots at the $\sim$ 1 kpc level of an extended radio galaxy \citep[see][for one example]{mahony2022}, however this is a more complicated situation where individual follow-up is likely to be needed in order to identify the intervening galaxy. 

\begin{table*}[ht!]
    \centering
\caption{Overview of flux-limited radio surveys with near-complete spectroscopic redshift information.}
    \begin{tabular}{lrrrrrrr}
\hline
      Survey   &  3CR & MRC-5Jy & MRC-1Jy & CENSORS &  \\
\hline
      Number of sources &  173 & 174 & 558 & 150 &  \\
      Radio frequency   & 178\,MHz  & 408\,MHz & 408\,MHz & 1.4\,GHz \\
      Flux density limit & 10\,Jy & 5\,Jy & 1\,Jy & 7.5\,mJy \\
      Redshift completeness & 96\% & 98\% & 68\% & 71\% \\
      Median redshift & 0.48 & 0.47 & 0.69 & 1.01  \\
References & \cite{laing1983} & \cite{best1999} & \cite{McCarthy1996} & \cite{brookes2008} \\
\hline
    \end{tabular}
    \label{tab:stats}
\end{table*}

\subsection{The expected radio-source populations}
Based on earlier studies \citep[e.g.][]{Best2012,Heckman2014,Ching2017}, we can expect the FLASH radio sources with $S_{856} >$ 30\,mJy to fall into three main classes: 
\begin{itemize} 
\item 
{\bf Low-excitation radio galaxies} (LERGs) are the dominant radio AGN population out to at least $z\sim0.8$ \citep{Ching2017}. These are massive galaxies, but lack a radiatively efficient accretion disk so their optical spectra show weak or no emission lines \citep{Heckman2014}. 
\item 
{\bf High-excitation radio galaxies} (HERGs) have a thin, radiatively-efficient accretion disk surrounding the central black hole. Their optical spectra show strong, narrow emission lines. 
\item 
{\bf Radio-loud quasars} also have a radiatively-efficient accretion disk, and their optical spectra show both bright AGN continuum radiation and strong, broad emission lines. HERGs and quasars are postulated to be the same class of objects seen at different orientations, with the central AGN in HERGs largely blocked in the optical by a dusty torus \citep[e.g.][]{urry1995}. 
\end{itemize}

\noindent Two other radio-source populations, `normal' inactive galaxies \citep{Condon_1992} and low-luminosity or radio-quiet (RQ) AGN \citep{Panessa_2019}, 
for which the origin of the radio emission is hotly debated and, at least in some cases, arises from something other than processes related to star formation (SF), start to emerge in radio surveys at flux densities below 1-10\,mJy but are not expected to occur in significant numbers in the FLASH sample. The relative fractions of HERGs, LERGs, and quasars within the FLASH sample remain uncertain, as these populations have only been systematically characterised out to $z\sim0.7$ through cross-matching NRAO VLA Sky Survey \citep[NVSS,][]{Condon_1998} and Faint Images of the Radio Sky at Twenty Centimeters \citep[FIRST,][]{Becker_1995} survey with wide-area optical redshift surveys \citep[e.g.,][]{Best_2005,Sadler_2007,Ching2017}. At higher redshifts ($z>0.7$) their demographics are less well defined, since the host galaxies of most LERGs, become too faint for reliable optical redshifts to be measurable with 4\,m-class telescopes used for existing large multi-object redshift surveys. Nevertheless, existing studies show that HERGs and LERGs exhibit distinct evolutionary trends with redshift \citep[][]{pracy2016,kondapally2023}. In addition, photo-$z$s for LERGs are generally reliable out to at least $z\sim1$ \citep{Ching2017}. Since the authors show that the three main radio-source populations (LERGs, HERGs, and quasars) can be distinguished through their optical fluxes as a function of redshift\footnote{$g-i$ colour at $z>0.4$}, it may be possible to map out the redshift distribution of these three populations further, including fainter sources,  using photo-$z$s and optical photometry alone. This could be a key aspect in investigating the rapid decline of LERGs beyond $z\sim 1$, possibly reflecting genuine evolutionary changes in the galaxy population, incompleteness due to survey depth limits, or the lack of appropriate training sets for photo-$z$ estimation of these objects.

\subsection{The expected $n(z)$ of FLASH continuum sources}

In the FLASH survey, almost all the background radio continuum sources used to search for \hi absorption will be distant radio galaxies that lack an optical redshift. This is noticeably different from large optical surveys searching for Damped Lyman-alpha (DLA) absorbers, where the redshifts of the background quasars are commonly already known \citep[][]{Srianand_2005, Ellison_2008, SR_2016, Hu_2023}.

\Cref{tab:stats} gives some guidance as to the likely redshift distribution of ASKAP FLASH radio sources with $S_{856} >$ 30\,mJy. Sources in the northern 3CR \citep[][]{laing1983} and southern MRC-5Jy \citep[][]{Best_2005} surveys, with a flux-density limit of several Jy and thus biased toward lower redshifts, have a median redshift of $z\sim0.5$, but this increases to medians of $z\sim0.7$ and $z\sim1.0$ for the MRC-1Jy \citep[][]{McCarthy1996} and the much fainter CENSORS \citep{brookes2008} sample, respectively \citep[][]{dezotti2010}. Together, these suggest a likely median redshift in the range $z\sim0.8-1$ for the ASKAP FLASH continuum sources above 30\,mJy. That said, the median $z$ of RL AGN from both modelling \citep[SKADS, T-RECS, e.g.][]{Raccanelli_2015} and photo-$z$ \citep[e.g.][]{Luken_2023} is expected to be $\gtrsim 1$. \cite{hardcastle2025} recently published a catalogue of photo-$z$s for around 600\,000 radio AGN candidates brighter than 1.1\,mJy in the LOFAR Two-Metre Sky Survey \citep[LoTSS,][]{Shimwell_2017}. Limiting their catalogue to a subset of sources with  $S_{144 \, \rm MHz} >\, $30\,mJy$ \, \bigl (\frac{144}{856}\bigl )^{- \alpha}$, assuming a radio spectral index $\alpha = 0.7$, and further restricting it to single-component sources (S\_code=S), we find a median redshift of $z=1.00$, very close to the CENSORS value.

\section{The {\sc{PICZL}} code}
\label{sec:piczl}

{\sc{PICZL}}\footnote{\url{https://github.com/williamroster/PICZL}}, developed by \cite{roster2024}, is a deep learning framework designed to address the challenge of reliable photo-$z$ estimation for extragalactic sources with complex emission budgets, particularly AGN \citep[][]{Salvato2011, Brescia_2019, Saxena2024}. Working directly with pixel-level flux distributions \citep[][]{Hoyle_2016, Pasquet_2018, Hayat_2021, Newman_2022}, it integrates convolutional neural networks (CNNs) with broad-band optical and mid-IR imaging from the Dark Energy Spectroscopic Instrument (DESI) Legacy Imaging Surveys Data Release 10 \citep[LS10,][and the DECam eROSITA Survey (DeROSITAS; PI: Zenteno)]{Dey2019}. LS10 includes data from the Dark Energy Survey \citep[DES,][]{Abbott2016}, Beijing-Arizona Sky Survey \citep[BASS,][]{Zou2017}, Mayall z-band Legacy Survey \citep[MzLS,][]{Silva2016}, and bands W[1,2,3,4] from the Near-Earth Object Wide-field Infrared Survey Explorer \citep[NEOWISE,][]{Mainzer2011,Lang_2014,Meisner2017}. The use of spatially resolved colour information constructed through band ratios, allows the model to learn meaningful morphological and spectral patterns without relying on manually engineered features, such as template or model fluxes, which often do not account for an AGN component when derived \citep[see e.g.][]{collister04, carrasco13, hogan15, almosallam16, Disanto_2018}. This includes learning the importance of redshift-dependent light profiles, through which {\sc{PICZL}} gains a more nuanced understanding of source type and redshift, particularly where AGN contamination or morphological diversity complicate interpretation.

\subsection{Generic extragalactic populations beyond AGN}

The versatility of {\sc{PICZL}} lies in its adaptability. It can be retrained or fine-tuned for different source populations, while its architecture supports scalable deployment across large imaging datasets, reducing reliance on deep spectroscopy or handcrafted spectral-energy distribution (SED) models. Consequently, while {\sc{PICZL}} was originally developed and optimized for optically and Xray selected AGN across the interval $0 \leq z \leq 8$, it is also effective when applied to inactive galaxies predominantly identified at lower redshifts. When trained on such a sample, {\sc{PICZL}} achieves redshift estimation performance on par with much deeper, pencil-beam surveys that rely on extensive multiband photometry for template fitting (Götzenberger et al. in prep.). This result highlights the model’s ability to extract redshift-sensitive features from shallower, wide-field imaging data, making it well-suited for current and upcoming large-scale surveys like the Vera C. Rubin Observatory Legacy Survey of Space and Time \citep[LSST,][]{Ivezic2019}, and \textit{Euclid} \citep[][]{Mellier2025}.

\begin{figure*}[t!]
\centering
\includegraphics[angle=0,width=1.0\hsize]{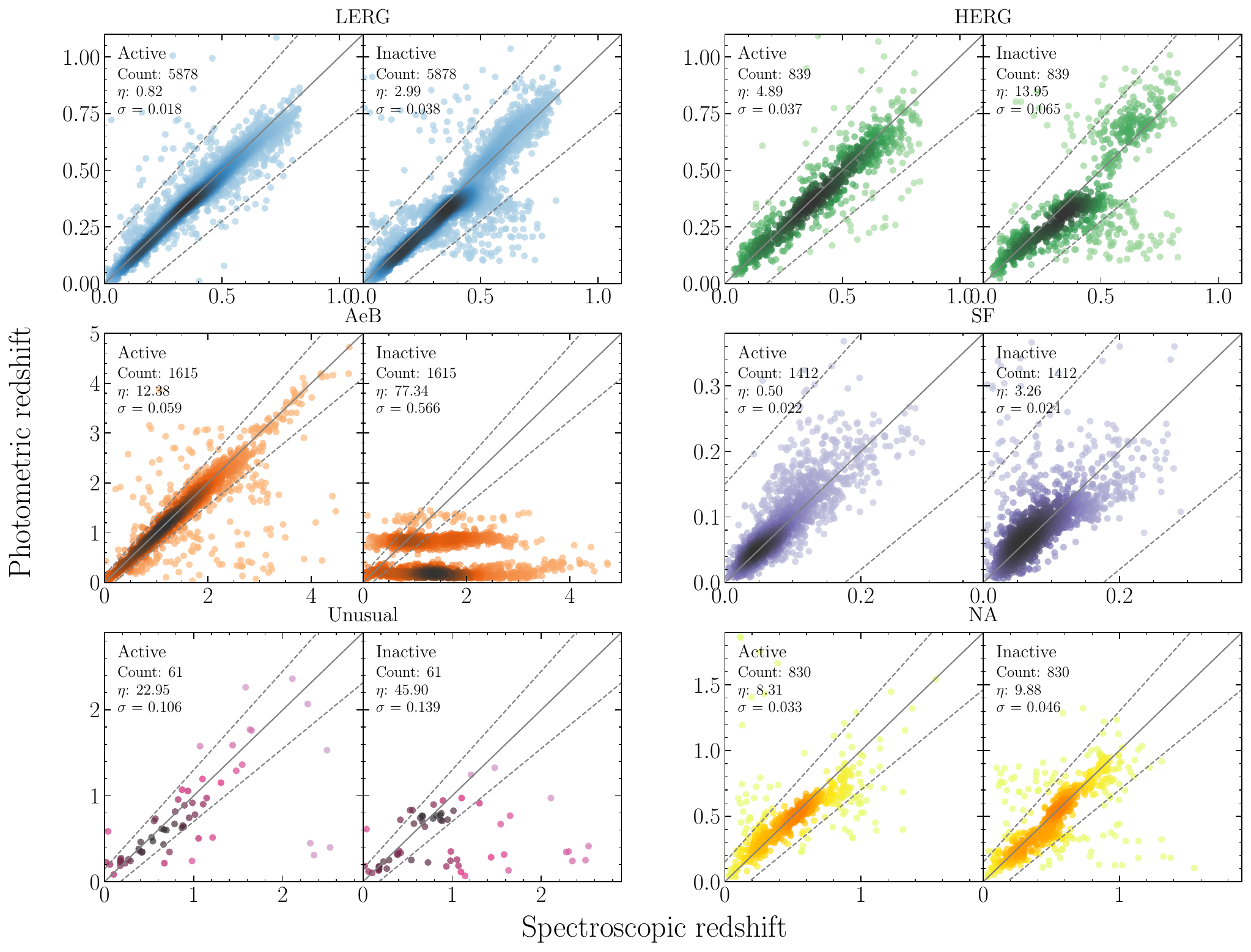}
\caption{Photo-$z$ versus spec-$z$ for six distinct spectroscopic subclasses of radio galaxies (HERG, LERG, AeB, SF, Unusual, and NA) following the classification scheme of \cite{Ching2017}}. Each panel displays the photo-$z$ performance for different source types across redshift, with subpanels separated by {\sc{PICZL}} mode (Active, Inactive). The identity line and the outlier boundary condition (dashed) are shown in grey for reference. For each subclass, the number of sources, accuracy $\sigma_{\rm NMAD}$ and fraction of outliers $\eta$ are reported. All scatter plots are colour-coded by their respective kernel density.
\label{fig:1}
\end{figure*}

\subsection{{\sc{PICZL}} uncertainty}

To capture redshift uncertainty, {\sc{PICZL}} includes a Gaussian mixture model (GMM) backend that transforms the network's outputs into full probability density functions (PDFs). This probabilistic framework allows for robust treatment of multimodal distributions, indicative of ambiguous cases, and facilitates more scientifically rigorous downstream analyses. Additionally, ensemble learning is used to combine predictions from multiple independently trained models, improving generalisation and reducing susceptibility to outliers.

\section{Application of PICZL to radio-selected samples}
\label{sec:tests}

We explore {\sc{PICZL}}’s performance beyond its native tracer domain, by applying the model to radio-selected samples, since approximately half of the sources in modern radio surveys are known to host an AGN \citep{Norris2013}, the vast majority of can only be detected in the X-rays when observed at very deep sensitivities \citep{Lafrance2012, Smolcic2017}. 

\subsection{Covariate shift biases}
\label{CS}

Applying {\sc{PICZL}} to source populations selected differently from those comprising its training set can introduce systematic biases rooted in both varying galaxy properties and survey characteristics \citep[][]{Duncan2018a,Duncan2018b,norris2019,Treyer_2025}. For instance, AGN selected via radio or X-ray emission may exhibit markedly different SEDs even at identical redshifts, possibly leading to divergent optical appearances and subsequent colour-redshift degeneracies. The observed populations are further modulated by survey depth, where deep surveys preferentially detect fainter, more weakly emitting AGN, while shallower ones capture the brighter, more extreme population \citep[][]{Salvato2011,Hsu2014}. As a result, these selection effects shape the $n(z)$ of the source population, which in turn sets the statistical priors that the model implicitly learns during training. Consequently, when {\sc{PICZL}} is applied to a previously unseen dataset of distinct underlying true $n(z)$, these inherited priors can skew the resulting photo-$z$ posteriors, leading to population-dependent biases in the predictions.

\subsection{Performance verification}
\label{PV}
 
To ensure that {\sc{PICZL}} yields reliable redshift estimates when applied to radio-selected FLASH continuum sources, we assess its performance on three independent test samples of various sources with good spectroscopic completeness that reflect the diversity of radio-emitting galaxy populations. These include: \cite{Best2012}, composed of HERGs, LERGs, and SFGs identified in the Sloan Digital Sky Survey \citep[SDSS,][]{York2000} up to $z \lesssim 0.7$; \cite{Ching2017} with WiggleZ and Galaxy And Mass Assembly (GAMA) spectroscopy covering both HERG/LERG galaxies up to $z \sim 0.7$ and quasars to extend the redshift baseline beyond $z > 2$; as well as the Molonglo Reference Catalog/1 Jansky Radio Source Survey \citep[MRC-1Jy,][]{McCarthy1996,Kapahi1998a,Kapahi1998b,Baker1999}, representing powerful RL AGN and quasars reaching $z \sim 2$. These datasets offer a controlled setting to explore the potential biases discussed above. By benchmarking {\sc{PICZL}} against these samples, we aim to build confidence in its applicability to FLASH continuum sources, many of which are expected to be massive, passive galaxies or moderate AGN that also emit in the X-rays or mid-IR (MIR).

To assess the performance of our photo-$z$ estimates, we adopt two widely used evaluation metrics \citep[][]{Ilbert_2006, Luken_2023}. First, we quantify the overall accuracy $\sigma_{\rm{NMAD}}$ as a robust estimate of the scatter between photo-$z$ and spec-$z$, calculated as
\begin{equation}
\sigma_{\textrm{NMAD}} = 1.4826\,  \textrm{median} \left[ \frac{|z_{\textrm{phot}} - z_{\textrm{spec}}|}{1 + z_{\textrm{spec}}} \right] \, .
\end{equation}
Second, we compute the outlier fraction, $\eta$, which captures the fraction of cases where photo-$z$s diverge significantly from spec-$z$s. An object is classified as an outlier if
\begin{equation}
\eta =
\frac{|z_{\textrm{phot}} - z_{\textrm{spec}}|}{1 + z_{\textrm{spec}}} > 0.15 \, .
\end{equation}

\Cref{fig:1} displays photometric versus spectroscopic redshift scatter plots for six spectroscopic subclasses of radio AGN from \cite{Ching2017}, where each panel is split into active\footnote{trained on optically and X-ray selected AGN up to $z \sim 6$, see Figure 2 in \cite{roster2024}} and inactive\footnote{trained on DESI redshifts of inactive galaxies up to $z \sim 1$ in the COSMOS field \citep[][]{Scoville_2007}} mode predictions from {\sc{PICZL}}. These sources are generally bright in LS10 photometry, meaning their photo-$z$ accuracy is overall not strongly limited by photometric uncertainties. This provides a relatively clean setting to assess {\sc{PICZL}}’s performance on a radio-selected AGN population, despite the fact that the model was not trained on radio galaxies specifically. Overall, results show robust redshift recovery, with low outlier fractions and high accuracy across all subclasses. In particular, sources classified as LERGs, typically associated with radiatively inefficient accretion and weak optical AGN features, display lower $\eta$ and $\sigma$ than HERGs or broad-line AGN (AeB), which exhibit strong nuclear continuum. Notably, sources labelled as ‘Unusual’ or unclassified (‘NA’), which often present challenges for template-based methods due to irregular or poorly understood SEDs, are nonetheless well handled by {\sc{PICZL}}. Interestingly, the active-mode predictions consistently outperform the inactive-mode ones across all classes. This is especially evident for AeB sources, which are more likely to appear as AGN at all wavelengths. This is an intriguing result given that most radio AGN (e.g. LERGs) have been known to appear as weak emitters in both optical and X-ray. Therefore, one might have expected the inactive model to perform better. However, AeB objects are characterised by a significant AGN contribution to their optical continuum in addition to broad emission lines, such that their broadband SEDs deviate strongly from those of normal galaxies, making galaxy-based templates or models unsuitable. The distinctive horizontal banding of AeB predictions from the inactive model can be understood as a combination of training-set limitations and feature-driven degeneracies. Since the model is trained exclusively on inactive galaxies at $z \lesssim $, sources at higher redshift lie outside the training regime, causing the model to preferentially map them onto redshift ranges that are well represented in the training data. At the same time, the predictions cluster around specific redshifts ($z \sim 0.4$ and $z \sim 0.8$) where, in galaxies, strong spectral features such as the Ca II 4000\si{\angstrom} break or prominent emission lines align with the LS10 bands and provide robust constraints. When applied to the more featureless, power-law–dominated AGN spectra, the model nevertheless anchors predictions to these “high-confidence” regions in colour–redshift space, leading to degeneracies and the observed horizontal stripes. The results of the remaining spectroscopic classes suggest that radio galaxies with weak optical features or a lack of detected X-ray emission in this work retain sufficient characteristic signatures in their SEDs for more accurate estimates by the AGN-trained model. These features might include subtle emission lines, mid-IR excess, or colour signatures linked to AGN activity that distinguish them from purely inactive galaxies. Consequently, the inactive mode predictions often  overestimate redshifts in low-$z$ star-forming systems, while the active mode maintains better fidelity overall. This exercise highlights the value of training ML techniques on a broad range of AGN populations, where the use of multiwavelength selection is not a drawback but a strength as each band, sensitive to different physical processes, recovers distinct subpopulations. Harnessing this complementarity will be an important aspect of future work in this field. As a consequence of the results presented in this section, we adopt {\sc{PICZL}} in active mode with no refinements made to the previously published version of the model to compute all photo-$z$ estimates presented in this work. 

\begin{figure}[t!]
\centering
\includegraphics[angle=0,width=1.0\hsize]{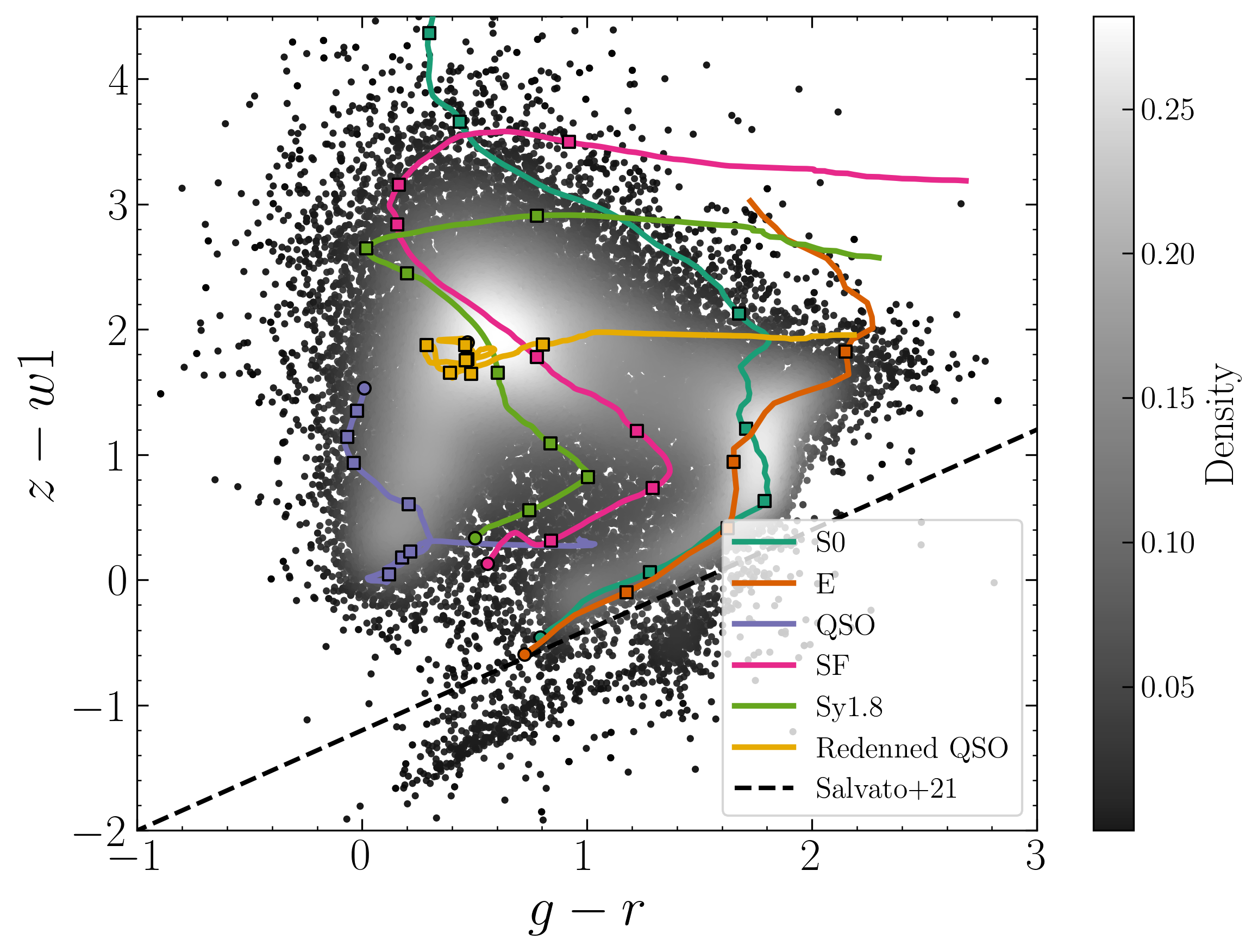}
\caption{Colour–colour diagram showing $g-r$ vs. $z-W1$ for sources with S/N $\geq 3$ in $g, r, z$, and $W1$. Points are colour-coded by Gaussian kernel density. Overplotted are empirical template tracks of various galaxy types, with square markers denoting specific steps in redshift (0.2, 0.4, 0.6, 1, 1.6, 2, 3) and circles marking the starting redshift. The black dashed line represents the \citep[][]{Salvato2022} Galactic/extragalactic selection boundary in colour space.}
\label{fig:4}
\end{figure}

\section{Compact-source sample}
\label{sec:CSS}

At the time of this analysis, approximately one-quarter of the full FLASH survey footprint has been observed and processed (refer to \Cref{fig:intro}). By combining the component catalogues from all available fields, we begin with a parent catalogue comprising roughly 2.3 million radio entries. Prior to being able to compute reliable photo-$z$, we first need to identify the optical CTPs of all FLASH continuum sources.

\subsection{Sample downselection}

To select a subset of FLASH-detected sources likely to provide sufficient continuum signal for \hi absorption measurements, we apply a flux-density threshold of $S_{856} > 30\,\rm{mJy}$. This reduces the sample to approximately 71\,000 entries. Since individual radio sources in the FLASH island catalogue can be resolved into multiple components, we restrict our analysis to islands made up of a single component. Although approximately 90\% of FLASH continuum sources are expected to be compact and unresolved in ASKAP images, this step thus reduces ambiguous associations during cross-matching with LS10 to identify the most probable multi-wavelength CTP. This procedure, including the use of a radio prior to and the validation of candidate matches, is described in detail in \ref{app:1}. Here, we focus on the resulting catalogue of compact FLASH continuum sources with reliable associations, which forms the basis of the analysis presented in the following sections.

\subsection{Galactic and extragalactic classification}
We examine the multiwavelength characteristics of the CTP sample beginning with an assessment of the extragalactic content. Adopting the definition of \cite{Salvato2022} based on optical LS10 colours, we find that 97\% of the sources appear extragalactic (i.e. above the dashed line). The resulting distribution, shown in \Cref{fig:4} and colour-coded by kernel density, reveals a region of high density associated with elliptical and luminous red galaxies (LRGs). Beyond the dominant AGN population, a subset of sources occupies regions of colour–colour space commonly associated with reddened quasars, Seyfert galaxies, and SFGs. While the colour distribution alone does not allow a quantitative decomposition of the underlying source classes, independent analysis demonstrates that the contribution from SFGs is small. A detailed assessment of the AGN and SFG content of the sample is therefore deferred to \ref{app:2}. For the remainder of this work, we restrict all figures and statistical analyses to the extragalactic sample, while retaining Galactic sources in the released catalogue.

\begin{figure}[t!]
\centering
\includegraphics[angle=0,width=1.0\hsize]{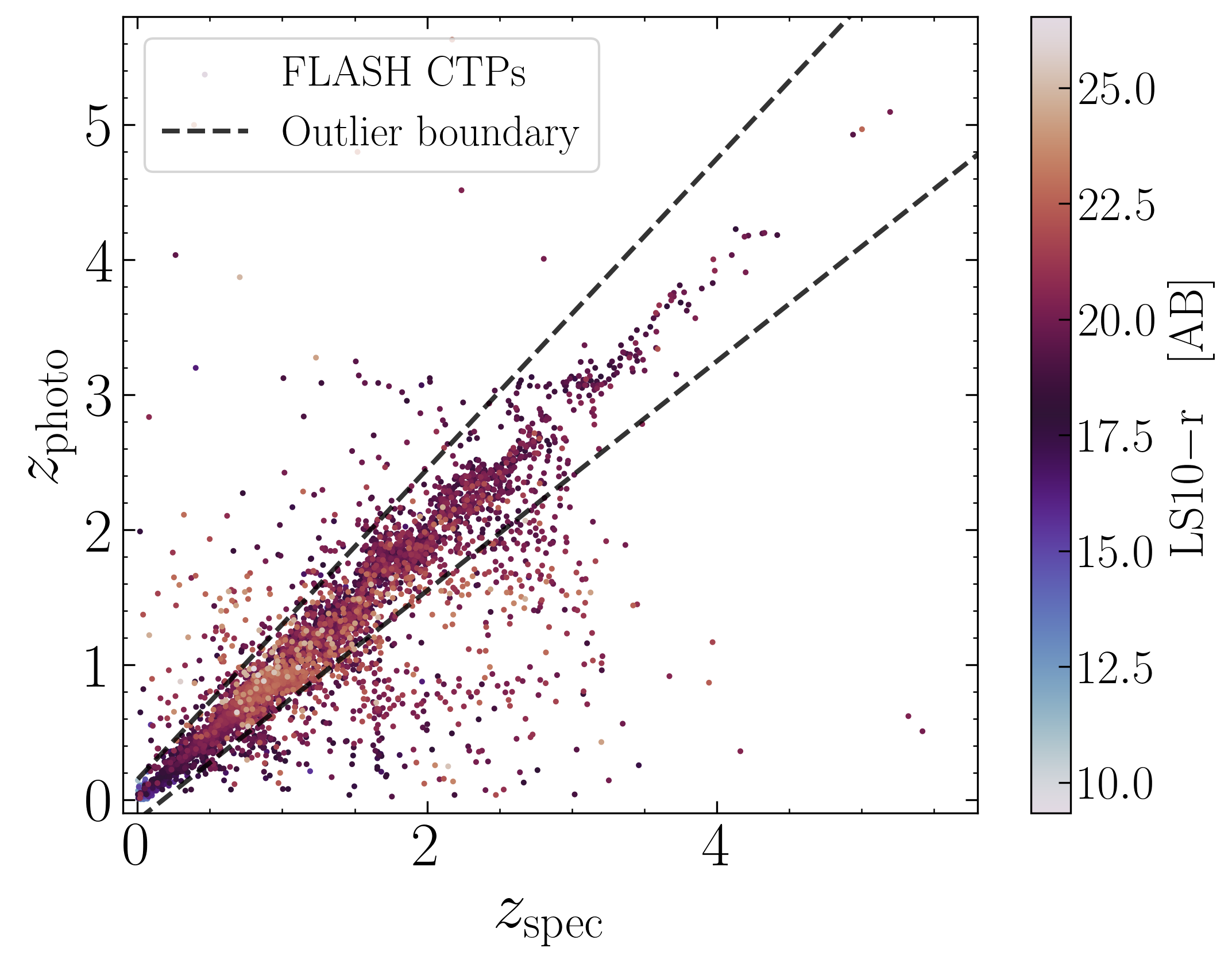}
\caption{Spectroscopic versus photometric redshifts for high-$z$ CTPs in the FLASH sample. Points are colour-coded by and plotted in increasing order of their LS10-$r$ band AB magnitude. The dashed lines indicate the 
$\frac{|\Delta z|}{1+z} \geq 0.15$ boundaries used to identify outliers.}
\label{fig:highz}
\end{figure}

\begin{figure}[t!]
\centering
\includegraphics[angle=0,width=1.0\hsize]{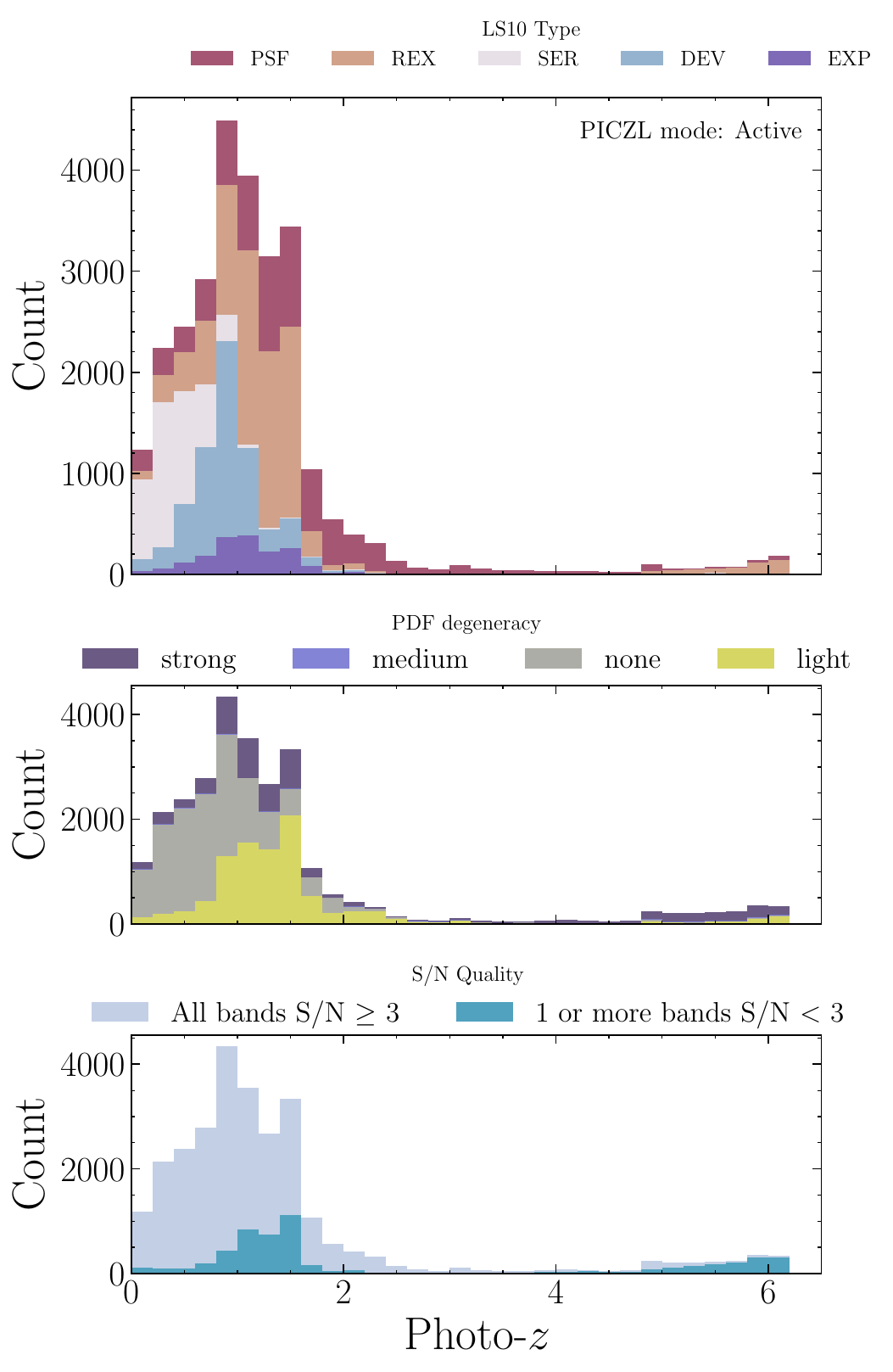}
\caption{\emph{Top}: FLASH compact source {\sc{PICZL}} photo-$z$ distributions coloured by different LS10 morphological classes. \emph{Middle}: Same distribution as above but coloured by the PDF degeneracy class: none — single strong peak; light — weak secondary peak(s); mild — noticeable but subdominant secondary peaks; strong — multiple peaks of comparable strength. \emph{Bottom}: Same distribution as above but coloured by the S/N of all LS10 ($g,r,i,z$) bands.}
\label{fig:2}
\end{figure}

\section{Photometric redshift estimates}
\label{sec:red_results}

Photo-$z$s for all identified CTPs are estimated using {{\sc PICZL}}, based on LS10 imaging and adopting the configuration described in \Cref{PV}. Roughly 13\% of these can successfully be matched within 1 arcsec to a compilation of public spec-$z$s \citep[][Igo et al. in prep.]{Kluge_2024}. As displayed in \Cref{fig:highz} our photo-$z$s perform robustly across this sample at $\eta = 12.37\%$ and $\sigma = 0.055$, showing no significant degradation even for fainter sources. The resulting redshift distribution, $n(z)$, is shown in \Cref{fig:2}. We find that approximately 13\% of continuum sources lie at $z<0.42$ (foreground), 35\% within the detectability range of FLASH (`in-band'), and 52\% at $z>1$ (background), indicating that more than half of the continuum sources cannot be searched for associated \hi absorption. Consequently, the effective sample available for associated absorption systems is significantly reduced, increasing the relative likelihood of detecting intervening \hi absorption along the line of sight. The low-redshift regime appears dominated by sources best fit by a Sérsic (Type = SER) profile, while de Vaucouleurs (Type = DEV) profiles peak around $ z\sim 1$, with Type = REX and Type = PSF sources prevailing at higher redshifts. In particular, the drop in sources beyond $z \gtrsim 1.5$ reflects the optical emission shift into the UV rest-frame, which can be attenuated by dust or simply intrinsically weak, where they may appear faint enough to fall below the detection threshold of LS10. Consequently, we presume much of the high-redshift excess at $z \gtrsim 5$ to arise from the prior imprint of the AGN training sample (see \Cref{CS}). However, while the overall sample shows relatively well-constrained estimates, with a median 1$\sigma$ error of 0.67, most of these sources are associated with strongly degenerate PDFs, often featuring a secondary PDF peak at a more plausible redshift. Nevertheless, they represent a subset of potentially interesting cases for further targeted investigation. The median redshifts by morphological class, and for the total sample, are given in \Cref{tab:redshifts}, including and excluding ($z \leq 3$) considering the high-$z$ tail. Future spectroscopic follow-up will play an important role in expanding the sample of secure redshifts, providing an opportunity to cross-check the photo-$z$s presented here while also enhancing the overall redshift completeness.

\subsection{Connecting radio, optical and X-ray}
\label{sec:beyond}

\begin{figure}[t!]
\centering
\includegraphics[angle=0,width=1.0\hsize]{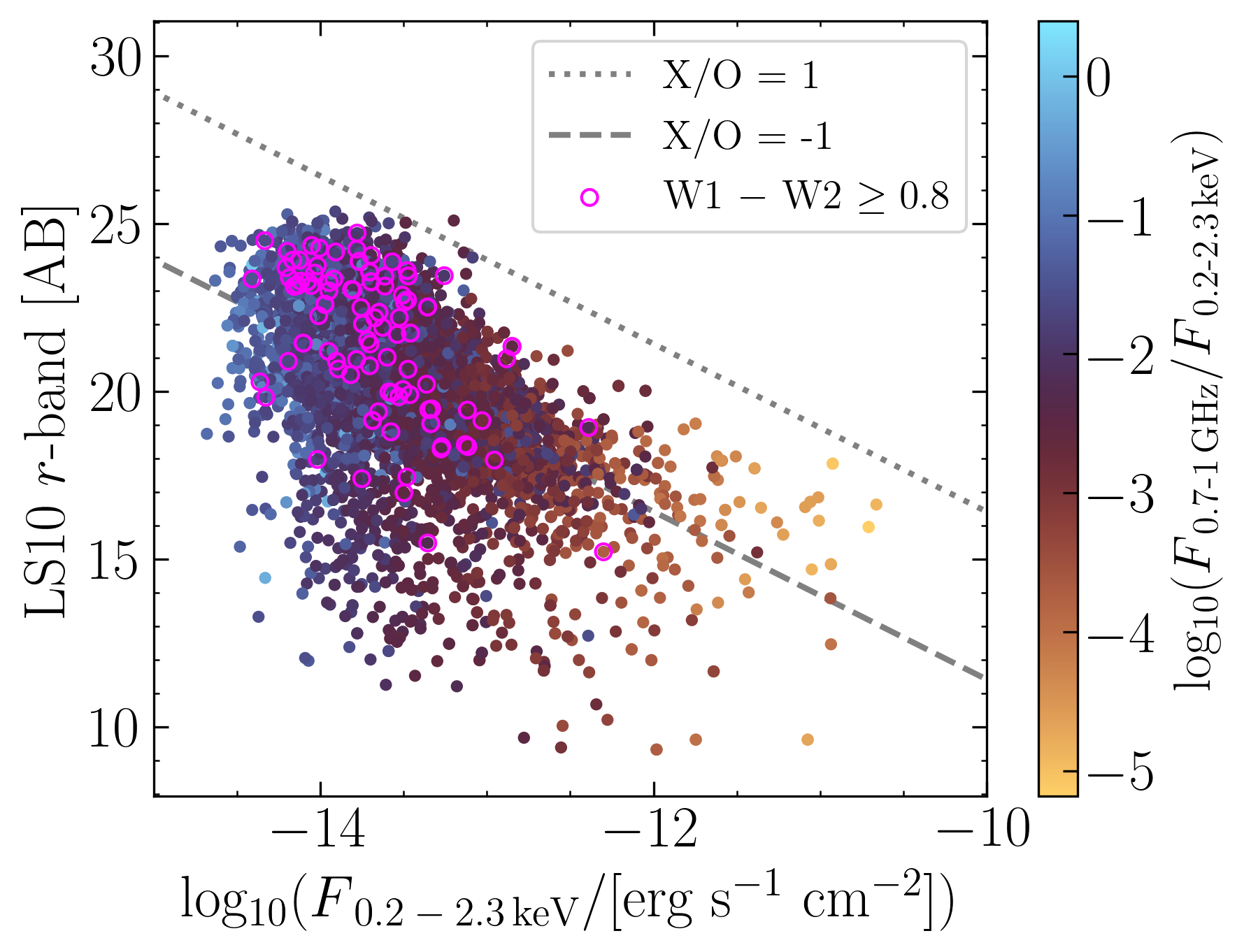}
\caption{Optical $r$-band magnitudes as a function of soft X-ray flux for FLASH CTPs successfully cross-matched to eRASS:5. Sources are colour-coded by their radio over X-ray (both in units of $\rm{erg}\,\rm{s}^{-1}$) flux ratios. Additionally, we highlight AGN selection according to \Cref{eq3} and |X/O| < 1 from \cite{Macca_1988}.}
\label{fig:9}
\end{figure}

Within the region of overlapping survey footprints, we identify approximately 6500 associations between the $\sim 22{,}000$ FLASH and the deepest stacked extended Roentgen Survey with an Imaging Telescope Array \citep[eROSITA,][]{Merloni2024} all sky survey (eRASS:5) LS10 CTPs. This $\sim 30\%$ detection fraction lies at the higher end of values reported in the literature, which generally range from 10–20\% for fainter or more extended radio populations \citep[e.g.][]{Lafranca_2010,DMoro_2013}. The majority of FLASH continuum sources, however, remain undetected in eRASS:5 as many compact, bright radio galaxies are presumably either jet-dominated systems with intrinsically weak X-ray cores, or, potentially obscured AGN whose soft X-ray emission falls below the eROSITA bandpass, highlighting both the diversity of the underlying AGN population and the selection biases introduced by soft X-ray surveys. As shown in \Cref{fig:9}, we observe objects from different populations: sources that are moderately bright in both X-ray and radio, and sources that are X-ray bright but comparatively radio faint. Although SF can contribute to radio and X-ray, both emission processes are ultimately linked to the accretion activity of the super-massive back hole (SMBH), with X-rays tracing the coronal emission near the black hole and radio tracing jet power \citep[][]{Igo_2024}. Noticeably, we find that a large fraction of the AGN in our sample populate a region characterised by relatively uniform radio-to–X-ray flux ratios at faint X-ray and optical flux levels. This regime also contains the majority of AGN selected via their IR colours, suggesting a population dominated by obscured and/or intrinsically X-ray–weak accretion. At the same time, our sample is biased toward luminous, accretion-powered systems by construction, and a substantially larger scatter in radio–X-ray behaviour is expected in deeper or more diverse samples. In particular, at intermediate luminosities, the coupling between radio and X-ray emission can be disrupted by a combination of nuclear obscuration, host-galaxy SF, and variations in jet production efficiency. Toward the brightest X-ray fluxes, the source density declines sharply, leaving only a small number of objects that preferentially exhibit lower radio-to–X-ray flux ratios, corresponding to X-ray–bright but radio–faint systems.

\begin{table}[t!]
    \centering
\caption{FLASH median redshifts per LS10 morphological class}
    \begin{tabular}{lrrrrr|r}
\hline
      Type   &  DEV & PSF & EXP & REX & SER & All \\
\hline
      w/ high-$z$ tail &  0.88 & 1.40 & 1.05 & 1.20 & 0.39 & 1.02 \\
      w/o high-$z$ tail   & 0.88 & 1.34 & 1.05 & 1.18 & 0.39 & 0.99 \\
\hline
    \end{tabular}
    \label{tab:redshifts}
\end{table}

\begin{figure}[t!]
\centering
\includegraphics[angle=0,width=1.0\hsize]{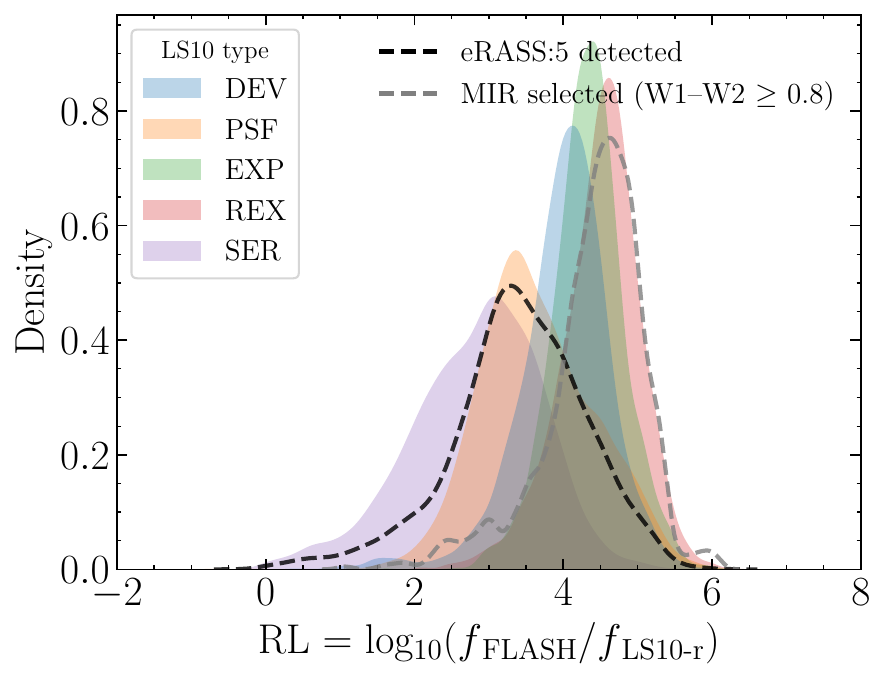}
\caption{Kernel density estimates of the logarithmic radio-to-optical loudness, $\log_{10}(f_{\rm{FLASH}}/f_{\rm{LS10}-r})$, for LS10 galaxies with reliable $r$-band detections (S/N $\geq$ 3). Filled curves show the distributions for different LS10 morphological types, while dashed lines indicate the subsets with X-ray detections in eRASS:5 (black) and MIR-selected ($W1 - W2 \geq 0.8$, grey). The KDEs are plotted with common\_norm=False, so the curves reflect the absolute fraction of the parent LS10 sample represented by each subset, rather than being rescaled to integrate to unity.}
\label{fig:6}
\end{figure}

\subsection{Infrared properties}

Next, we explore a relative measure of radio loudness (RL) of our sources as a function of their LS10 morphological classification. We define this quantity as the logarithmic ratio of radio flux density to optical flux using the available $\rm{LS10}\,r$-band and 1.4 GHz measurements scaled from FLASH fluxes (see \Cref{app:2}), which differs from the definition of radio loudness in the classical sense \citep[][]{Kellermann_2016}. Given these definitional differences, the RL values presented here should not be interpreted in terms of the conventional RL/RQ division. Instead, they are intended to provide a homogeneous, internally consistent metric for identifying relative trends within the FLASH population. The resulting distributions, shown in \Cref{fig:6}, span values of $-2 \lesssim \mathrm{RL} \lesssim 6$, with a global peak at RL $\sim 4.5$. A clear dependence on source morphology is visible. Sérsic-profile or disc-like sources dominate the lower end of the distribution, peaking at RL $\sim 3$, consistent with a population of modestly RL galaxies, potentially including SF systems and low-excitation AGN. Optically point-like sources (PSF) show a unimodal distribution centred around moderate RL, peaking closer to $\rm{RL} \sim 4$, suggesting that literature claims of RQ versus RL bimodality among quasars might be an artefact of observational biases \citep[e.g.][]{Kellermann1989, Sikora07, Mahony_2012}. The remaining morphological classes all display a single dominant peak at $\rm{RL} \sim 4.5$, characteristic of classical RL AGN. When considering sources with additional multi-wavelength AGN signatures, we find that X-ray–detected AGN peak at a lower loudness compared to IR-selected AGN. Using the AGN selection criterion
\begin{equation}
\label{eq3}
    \rm{W}1 - \rm{W}2 > 0.8 \, ,
\end{equation}
as introduced by \cite{Stern_2012} and \cite{Assef_2018}, $\sim$38\% of the 24\,776 CTPs with S/N$_{\rm{W}1\&\rm{W}2} \geq 3$ satisfy this condition. Roughly 20\% of objects classified as AGN according to \Cref{eq3} are also detected in eRASS:5, provided they share the same LS10 CTP. Consequently, this represents a lower limit, as it excludes sources whose X-ray and radio detections were assigned to different optical CTPs. This indicates a significant overlap between the IR-selected and X-ray detected AGN populations within the shared footprint of the CTP sample \citep[][]{Mendez_2013}. However, the contrast in peaks displayded in \Cref{fig:6} may indicate that eROSITA preferentially identifies moderately RL systems, tracing potentially more massive, quiescent galaxies, while IR selection isolates more obscured AGN, inherently missed in the X-ray selection, found to preferentially be RL \citep[][]{Best_2005,Smolcic17,Wang_2024}.

\begin{figure}[t!]
\centering
\includegraphics[angle=0,width=1.0\hsize]{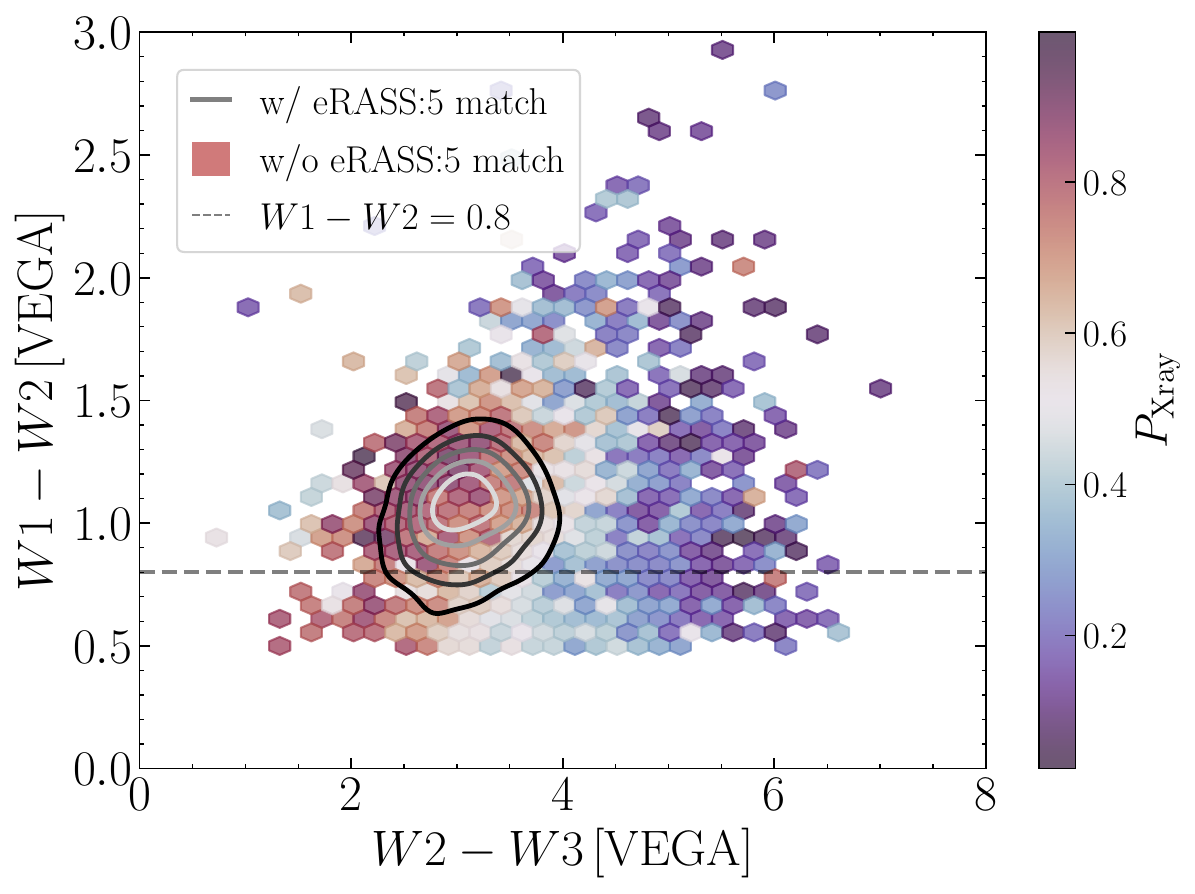}
\caption{WISE colour–colour diagram for sources with $W1 - W2 > 0.5$ using VEGA magnitudes. The background hexbin map represents the distribution of sources without an eRASS:5 match, coloured by the mean probability of being an X-ray emitter. Overplotted grey contours trace the kernel density of sources successfully matched to eRASS:5. The dashed horizontal line marks the reference threshold at $W1 - W2 = 0.8$.}
\label{fig:5}
\end{figure}

\subsection{Obscured sources}
\label{OS}
\begin{figure}[htbp]
\centering
\includegraphics[angle=0,width=1.0\hsize]{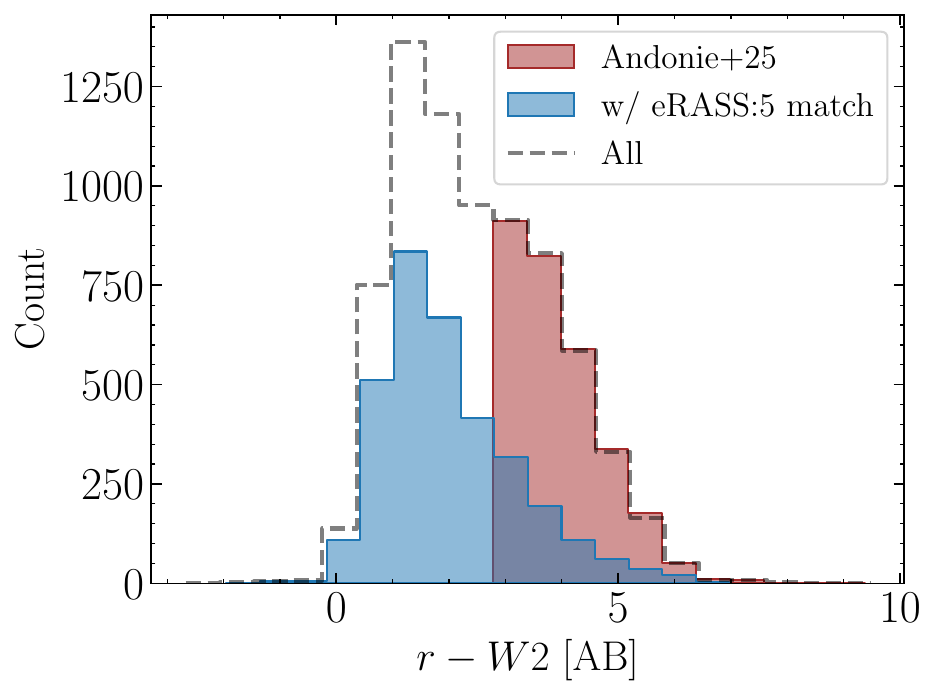}
\caption{FLASH CTP distribution in regard of the X-ray AGN obscuration selection as defined by \cite[][]{Hickox_2007}, showing the full sample (grey dashed line), those matched to eRASS:5 (blue), and the subset sources satisfying the obscuration threshold (red).}
\label{fig:5.1}
\end{figure}

Because eROSITA is most sensitive in the soft X-ray band, it naturally underdetects heavily obscured sources. This property allows us to use X-ray detections or the lack thereof as a diagnostic for obscuration within our sample. As shown in \Cref{fig:5}, FLASH radio sources successfully cross-matched with eRASS:5 largely fall into the region of colour space typically associated with a high probabilities of being an X-ray emitter, consistent with expectations. The respective probabilities are computed using a photometric prior defined by \cite{Salvato25}. In contrast, many sources without an X-ray counterpart cluster in the locus characteristic of obscured AGN \citep[][]{Wright_2010}. We further probe this obscured population using the optical–MIR colour diagnostic

\begin{equation}
    r - \rm{W}2 \geq 5.9\, [\rm{VEGA}]
\end{equation}
as defined by \cite{Hickox_2007} and utilised more recently in e.g. \cite{Andonie_2025} as the threshold for selecting mainly X-ray AGN with a hydrogen column density $N_{\rm{H}} \geq 10^{22}\, \rm{cm^{-2}}$ at low redshift, where these bands track the accretion disk and torus. \Cref{fig:5.1} shows that roughly 40\% of the sources displayed in \Cref{fig:5} pass this criterion. These include 23\% of the sources that were successfully cross-matched to eRASS:5. 

These results suggest that a substantial fraction of the FLASH sample occupies regions of colour space consistent with obscured AGN, underscoring the need for multiwavelength approaches to fully characterise AGN demographics.

\subsection{Environmental influence on radio power}
\label{sec:env}

\begin{figure}[t!]
\centering
\includegraphics[angle=0,width=1.0\hsize]{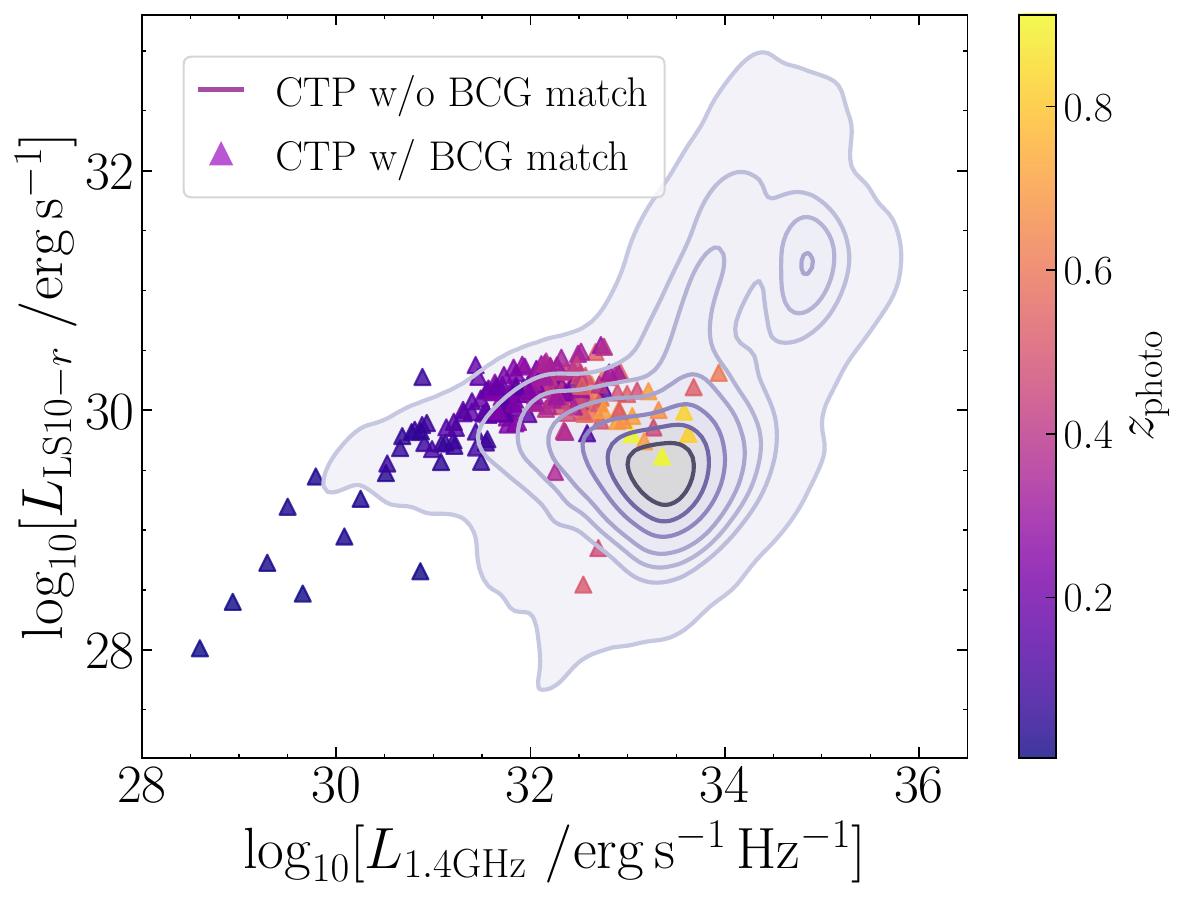}
\caption{Kernel density estimate of compact FLASH source luminosities, subject to {\sc {PICZL}} photo-$z$s, at 1.4 GHz and the LS10 $r$-band for sources matched to eROSITA extended clusters and non-BCG sources.}
\label{fig:8}
\end{figure}

To quantify the environmental influence of radio AGN, we investigate whether a subset of our radio sources can be associated with brightest cluster galaxies (BCGs) identified in eROSITA DR1 \citep{Kluge_2024,Balzer_2025,Veronica_25}.  BCGs are the most massive galaxies in clusters and are often found at or near the center of the gravitational potential well. We perform a positional cross-match between our radio sample and the BCG coordinates, requiring a maximum separation of 3 arcseconds. This radius is chosen 
as these kind of systems tend to be at low redshift where precise centering for extended galaxies is less accurate than for point sources. Because BCGs are known to frequently host powerful RL AGN, this provides a direct diagnostic of the impact of environment on AGN triggering \citep[][]{Shen_2020, Popesso_2024}. In particular, we compare the incidence of BCG-associated radio sources in clusters (122) against the remaining radio population as shown in \Cref{fig:8}. We find that the distribution of non-BCG galaxies peaks at higher radio luminosities but fainter optical magnitudes compared to cluster-associated sources, suggesting radio luminosities to be surpressed in dense cluster environments. Part of this trend is likely due to selection effects, such as the lower median redshift of the cluster sample as extended X-ray emission becomes increasingly difficult to detect at higher redshifts because of cosmological surface brightness dimming. In addition, our compact-source selection criterion likely biases against nearby extended, lobe-dominated systems, preferentially selecting core-dominated HERGs and compact LERGs. Nonetheless, genuine differences in AGN fueling channels (mergers and cold- flow-driven activity for high-$z$ powerful AGN vs. hot-halo/maintenance radio mode inside clusters) may also play a role \citep[][]{Best2012, hardcastle2025}.

\section{The role of photo-$z$s in identifying \hi absorbers}
\label{sec:improved_selection}

As a next step, beyond the scope of this paper, it will be important to examine the subset of radio sources exhibiting \hi absorption. This includes investigating whether these sources occupy distinct regions of optical, radio, or X-ray parameter space compared to sources without absorption, which could provide insight into the properties of AGN probed by associated absorbers. Confirming whether a detected absorption line arises from gas physically associated with the radio source (i.e. in the host galaxy) or from an intervening system along the line of sight currently requires tight (redshift) constraints. While spec-$z$s remain the gold standard, they are not always available, especially in large-area radio continuum surveys. Photo-$z$s provide a practical alternative by supplying a statistical estimate of the source redshift directly from multi-band optical and IR photometry. Even though photo-$z$s carry larger uncertainties compared to spectroscopy, they are sufficiently accurate to guide the identification and classification of \hi absorption features in several ways \citep[][]{Aditya_2024}:
\begin{itemize}
    \item Association with the background source: If the redshift of the presumed \hi absorption line coincides with the photo-$z$ of the optical CTP within the expected uncertainties, the absorption can be attributed to gas bound to the host galaxy.
    \item Identification of intervening absorbers: In cases where the absorption redshift is significantly lower than the photo-$z$ estimate of the optical background source, the line can be interpreted as an intervening absorber. This distinction allows us to probe the gaseous halos of galaxies along the line of sight not associated with the radio AGN host galaxy.
    \item Statistical population studies: Even when individual identifications are uncertain, photo-$z$s enable statistical analyses of large samples, such as the redshift evolution of associated versus intervening absorbers, or environmental dependencies. This includes investigating the total comoving absorption path length of the survey as well as estimating the spin temperature and hence interpreting the cold neutral medium fraction of absorbing galaxies\citep[][]{Allison_21, allison22}.
\end{itemize}

\noindent Ultimately, the goal of FLASH lies in investigating these individual classes in greater detail. Achieving this relies on reliably determining whether sources are physically connected or not. While photo-$z$s already provide some discriminatory power, recent promising approaches \citep[e.g.,][]{Curran_2021} have shown that these associations can be disentangled more effectively. In this context, photo-$z$ estimates could serve either as an additional ML input feature or as a tool to construct cleaner training samples. This, in turn, could pave the way toward future classification methods that require only basic spectral information, such as the width and depth of a single absorption line, to distinguish classes robustly.

\section{Outlook}
\label{sec:out}

As upcoming large \hi absorption surveys with instruments such as MeerKAT \citep[][]{Jonas_2016,Wagenveld24}, the Five hundred meter Aperture Spherical Telescope \citep[FAST,][]{Zhang25}, and the SKA \citep{Dewdney_2009} reach deeper flux density limits, they will open the possibility of probing higher-redshift radio sources, including quasars, down to flux densities of 1\,mJy. Already, using our FLASH sample, we tentatively select 1282 radio quasars in the {\sc{PICZL}} redshift range $2 \leq z \leq 4.5$ (see \Cref{fig:highz}). In this regime, spec-$z$ completeness reaches $\sim 50\%$ with consistent metrics as discussed in \Cref{sec:red_results}. Assuming the same survey area and selection efficiency as our FLASH sample, and scaling purely based on the flux density limit, we can estimate the increase in detectable sources using the Euclidean source count scaling 
\begin{equation}
    N(>S) \propto S^{\,-\frac{3}{2}}\ .
\end{equation}
This suggests, that SKA could detect roughly 160 times more high-$z$ radio quasars than currently found in our selection when lowering the current flux threshold of 30\,mJy down to 1\,mJy. We note that this is a conservative, first-order estimate and that more detailed forecasts should account for cosmological evolution of the luminosity function, K-corrections, and the flattening of radio source counts at sub-mJy fluxes.

\section{Summary}
\label{sec:summary}

In this work, we have demonstrated that {\sc{PICZL}}, which was developed for eROSITA (see \Cref{sec:piczl}), can produce accurate photo-$z$s for radio sources even without prior knowledge of their AGN nature (see \Cref{sec:tests}). We estimated redshifts for 45\,113 FLASH continuum sources using {\sc {PICZL}} applied to LS10 imaging, following the identification of their optical CTPs (see \Cref{sec:red_results} and \Cref{app:1}). To illustrate the scientific potential of the FLASH survey, we carried out a preliminary multiwavelength analysis, where, among other findings, we determined that roughly 30\% of FLASH sources are detected in the deepest eROSITA X-ray data, predominantly corresponding to objects that are bright in both X-ray and radio (see \Cref{sec:beyond}). A substantial fraction of sources, however, are radio-bright but X-ray faint, many of which are likely obscured AGN, as suggested by mid-infrared diagnostics (see \Cref{OS}). We further explored the potential impact of the cluster environment on the radio properties of FLASH sources ( see \Cref{sec:env}).

Looking forward, a key challenge for absorption-line science remains the reliable association of background sources with intervening absorbers (see \Cref{sec:improved_selection}). We outlined strategies to improve these classifications and provide a glimpse at the future of blind \hi absorption surveys (see \Cref{sec:out}). The expanding coverage of FLASH will soon enable more extensive redshift-based studies, unlocking a wide range of science opportunities from AGN demographics and multiwavelength source characterisation to detailed investigations of absorption-line systems.

\section{Data availability}
With this paper, we present a catalogue of photo-$z$s derived using {\sc PICZL} for the subset of FLASH continuum sources presented in this work that are cross-matched to LS10. A simplified description of the catalogue is given in \ref{app:3}. The full catalogue (including column description) is available at CDS via anonymous ftp to cdsarc.u-strasbg.fr (130.79.128.5) or via https://cdsarc.cds.unistra.fr/viz-bin/cat/J/other/PASA.

\begin{acknowledgement}

WR and MS acknowledge DLR support (Foerderkennzeichen 50002207). HY was supported by the National Research Foundation of Korea (NRF) grant funded by the Korea government (MSIT, RS-2025-00516062). RLD is supported by the Australian Research Council through the Discovery Early Career Researcher Award (DECRA) Fellowship DE240100136 funded by the Australian Government. This scientific work uses data obtained from Inyarrimanha Ilgari Bundara, the CSIRO Murchison Radio-astronomy Observatory. We acknowledge the Wajarri Yamaji People as the Traditional Owners and native title holders of the Observatory site. CSIRO’s ASKAP radio telescope is part of the Australia Telescope National Facility (\url{https://ror.org/05qajvd42}). Operation of ASKAP is funded by the Australian Government with support from the National Collaborative Research Infrastructure Strategy. ASKAP uses the resources of the Pawsey Supercomputing Research Centre. Establishment of ASKAP, Inyarrimanha Ilgari Bundara, the CSIRO Murchison Radio-astronomy Observatory and the Pawsey Supercomputing Research Centre are initiatives of the Australian Government, with support from the Government of Western Australia and the Science and Industry Endowment Fund.

This paper uses data that were obtained by The Legacy Surveys: the Dark Energy Camera Legacy Survey (DECaLS; NOAO Proposal ID\# 2014B-0404; PIs: David Schlegel and Arjun Dey), the Beijing-Arizona Sky Survey (BASS; NOAO Proposal ID\# 2015A-0801; PIs: Zhou Xu and Xiaohui Fan), and the Mayall z-band Legacy Survey (MzLS; NOAO Proposal ID\# 2016A-0453; PI: Arjun Dey). DECaLS, BASS and MzLS together include data obtained, respectively, at the Blanco telescope, Cerro Tololo Inter-American Observatory, National Optical Astronomy Observatory (NOAO); the Bok telescope, Steward Observatory, University of Arizona; and the Mayall telescope, Kitt Peak National Observatory, NOAO. NOAO is operated by the Association of Universities for Research in Astronomy (AURA) under a cooperative agreement with the National Science Foundation. Please see \url{http://legacysurvey.org} for details regarding the Legacy Surveys. BASS is a key project of the Telescope Access Program (TAP), which has been funded by the National Astronomical Observatories of China, the Chinese Academy of Sciences (the Strategic Priority Research Program``The Emergence of Cosmological Structures'' Grant No. XDB09000000), and the Special Fund for Astronomy from the Ministry of Finance. The BASS is also supported by the External Cooperation Program of Chinese Academy of Sciences (Grant No. 114A11KYSB20160057) and Chinese National Natural Science Foundation (Grant No. 11433005). The Legacy Surveys imaging of the DESI footprint is supported by the Director, Office of Science, Office of High Energy Physics of the U.S. Department of Energy under Contract No. DE-AC02-05CH1123, and by the National Energy Research Scientific Computing Center, a DOE Office of Science User Facility under the same contract; and by the U.S. National Science Foundation, Division of Astronomical Sciences under Contract No.AST-0950945 to the National Optical Astronomy Observatory. 

This work is based on data from eROSITA, the soft X-ray instrument aboard SRG, a joint Russian-German science mission supported by the Russian Space Agency (Roskosmos), in the interests of the Russian Academy of Sciences represented by its Space Research Institute (IKI), and the Deutsches Zentrum für Luft- und Raumfahrt (DLR). The SRG spacecraft was built by Lavochkin Association (NPOL) and its subcontractors and is operated by NPOL with support from the Max Planck Institute for Extraterrestrial Physics (MPE). The development and construction of the eROSITA X-ray instrument were led by MPE, with contributions from the Dr. Karl Remeis Observatory Bamberg \& ECAP (FAU Erlangen-Nuernberg), the University of Hamburg Observatory, the Leibniz Institute for Astrophysics Potsdam (AIP), and the Institute for Astronomy and Astrophysics of the University of T\"ubingen, with the support of DLR and the Max Planck Society. The Argelander Institute for Astronomy of the University of Bonn and the Ludwig Maximilians Universit\"at Munich also participated in the science preparation for eROSITA.

The Legacy Surveys consist of three individual and complementary projects: the Dark Energy Camera Legacy Survey (DECaLS; Proposal ID 2014B-0404; PIs: David Schlegel and Arjun Dey), the Beijing-Arizona Sky Survey (BASS; NOAO Prop. ID 2015A-0801; PIs: Zhou Xu and Xiaohui Fan), and the Mayall z-band Legacy Survey (MzLS; Prop. ID 2016A-0453; PI: Arjun Dey). DECaLS, BASS, and MzLS together include data obtained, respectively, at the Blanco telescope, Cerro Tololo Inter-American Observatory, NSF’s NOIRLab; the Bok telescope, Steward Observatory, University of Arizona; and the Mayall telescope, Kitt Peak National Observatory, NOIRLab. Pipeline processing and analyses of the data were supported by NOIRLab and the Lawrence Berkeley National Laboratory (LBNL). The Legacy Surveys project is honored to be permitted to conduct astronomical research on Iolkam Du’ag (Kitt Peak), a mountain with particular significance to the Tohono O’odham Nation.
NOIRLab is operated by the Association of Universities for Research in Astronomy (AURA) under a cooperative agreement with the National Science Foundation. LBNL is managed by the Regents of the University of California under contract to the U.S. Department of Energy.

This project used data obtained with the Dark Energy Camera (DECam), which was constructed by the Dark Energy Survey (DES) collaboration. Funding for the DES Projects has been provided by the U.S. Department of Energy, the U.S. National Science Foundation, the Ministry of Science and Education of Spain, the Science and Technology Facilities Council of the United Kingdom, the Higher Education Funding Council for England, the National Center for Supercomputing Applications at the University of Illinois at Urbana-Champaign, the Kavli Institute of Cosmological Physics at the University of Chicago, Center for Cosmology and Astro-Particle Physics at the Ohio State University, the Mitchell Institute for Fundamental Physics and Astronomy at Texas A \& M University, Financiadora de Estudos e Projetos, Fundacao Carlos Chagas Filho de Amparo, Financiadora de Estudos e Projetos, Fundacao Carlos Chagas Filho de Amparo a Pesquisa do Estado do Rio de Janeiro, Conselho Nacional de Desenvolvimento Cientifico e Tecnologico and the Ministerio da Ciencia, Tecnologia e Inovacao, the Deutsche Forschungsgemeinschaft and the Collaborating Institutions in the Dark Energy Survey. The Collaborating Institutions are Argonne National Laboratory, the University of California at Santa Cruz, the University of Cambridge, Centro de Investigaciones Energeticas, Medioambientales y Tecnologicas-Madrid, the University of Chicago, University College London, the DES-Brazil Consortium, the University of Edinburgh, the Eidgenossische Technische Hochschule (ETH) Zurich, Fermi National Accelerator Laboratory, the University of Illinois at Urbana-Champaign, the Institut de Ciencies de l’Espai (IEEC/CSIC), the Institut de Fisica d’Altes Energies, Lawrence Berkeley National Laboratory, the Ludwig Maximilians Universitat Munchen and the associated Excellence Cluster Universe, the University of Michigan, NSF’s NOIRLab, the University of Nottingham, the Ohio State University, the University of Pennsylvania, the University of Portsmouth, SLAC National Accelerator Laboratory, Stanford University, the University of Sussex, and Texas A\&M University.

BASS is a key project of the Telescope Access Program (TAP), which has been funded by the National Astronomical Observatories of China, the Chinese Academy of Sciences (the Strategic Priority Research Program “The Emergence of Cosmological Structures” Grant \# XDB09000000), and the Special Fund for Astronomy from the Ministry of Finance. The BASS is also supported by the External Cooperation Program of Chinese Academy of Sciences (Grant \# 114A11KYSB20160057), and Chinese National Natural Science Foundation (Grant \# 12120101003, \# 11433005). The Legacy Survey team uses data products from the Near-Earth Object Wide-field Infrared Survey Explorer (NEOWISE), a project of the Jet Propulsion Laboratory/California Institute of Technology. NEOWISE is funded by the National Aeronautics and Space Administration. The Legacy Surveys imaging of the DESI footprint is supported by the Director, Office of Science, Office of High Energy Physics of the U.S. Department of Energy under Contract No. DE-AC02-05CH1123, by the National Energy Research Scientific Computing Center, a DOE Office of Science User Facility under the same contract, and by the U.S. National Science Foundation, Division of Astronomical Sciences under Contract No. AST-0950945 to NOAO.

\end{acknowledgement}


\bibliography{refs}

\appendix

\section{Optical cross-matching}
\label{app:1}

We determine optical CTPs for FLASH continuum sources using the Bayesian cross-matching tool {\sc NWAY} \citep[][]{Salvato2018}, which combines astrometric information with photometric priors to improve the reliability of associations. In {\sc NWAY}, such priors are typically constructed by training a random forest model to distinguish the target source population, e.g. radio emitting objects, from the general field population \citep[][]{Salvato2022, Roster2025, Salvato25}. Because the FLASH sample is currently too small to construct a robust training set of its own, we adopt an existing radio prior built from secure, optically matched sources in the Rapid ASKAP Continuum Survey \citep[RACS,][]{McConnell_2020,Hale_2021}, based on the RACS-low observations at $S = 888 - 943.5\,\mathrm{MHz}$, approximately matching the observing frequency of FLASH. Using this prior, we run {\sc NWAY} on all selected FLASH continuum sources and identify, within 10 arcseconds, their most probable optical CTPs in LS10 \citep[][]{Salvato25}. To validate this approach, we compare the FLASH-based CTPs with those obtained by running {\sc NWAY} on RACS sources in the area jointly covered by LS10 and the footprint of eROSITA. For sources where the FLASH and RACS positions agree within 3 \,arcseconds (> 99\%), we find a CTP match rate exceeding 90\%, which demonstrates the reliability of the adopted prior. The remaining mismatches are most likely attributable to the substantially larger matching radius (1 arcmin) used in the RACS-based {\sc NWAY} associations.

The final catalogue comprises 45\,113 unique FLASH continuum sources with optical CTPs in LS10. Among these, 6305 cases correspond to instances where multiple FLASH entries (with distinct source IDs) are associated with the same optical CTP. This duplication arises primarily from the FLASH observing strategy, in which each field assigns unique identifiers independently. As a result, overlapping fields can register the same astrophysical source multiple times under different IDs. To mitigate this, we examined all multi-entry groups, where for groups of two sources, we retained a single entry when their angular separation was less than 5 arcseconds (accounting for 99\% of such cases). For groups with three or four members, all entries originated from distinct FLASH fields, and we likewise retained only one representative per group. This leaves us with a sample of $\sim$ 38\,600 unique FLASH CTPs. Approximately 2800 FLASH continuum sources remain without a CTP candidate entirely, $\sim 75\%$ of which lie within the Galactic plane where LS10 coverage is absent. The locations of the remaining sources are covered either by one (654) or all (473) Legacy bands, most of which are likely to be genuine `blank fields' where the optical CTPs are too faint to be visible in LS10 images, suggesting radio galaxies at $z \gtrsim 1$, as AGN-related processes (e.g. jets) can be radio-bright but optically/UV faint \citep[][]{Kondapally21}. In CENSORS, which has complete redshift coverage for a flux-limited radio sample ($> 7.5\,\textrm{mJy at}\,1.4\, \textrm{GHz}$), roughly 30\% of the sources have $z \geq 1.5$, and about 65\% of these have optical CTPs too faint to be detected in LS10. As a result, we expect $\sim 20\%$ of FLASH continuum sources to have no visible optical CTP in LS10.

\begin{figure}[t!]
\centering
\includegraphics[angle=0,width=1.0\hsize]{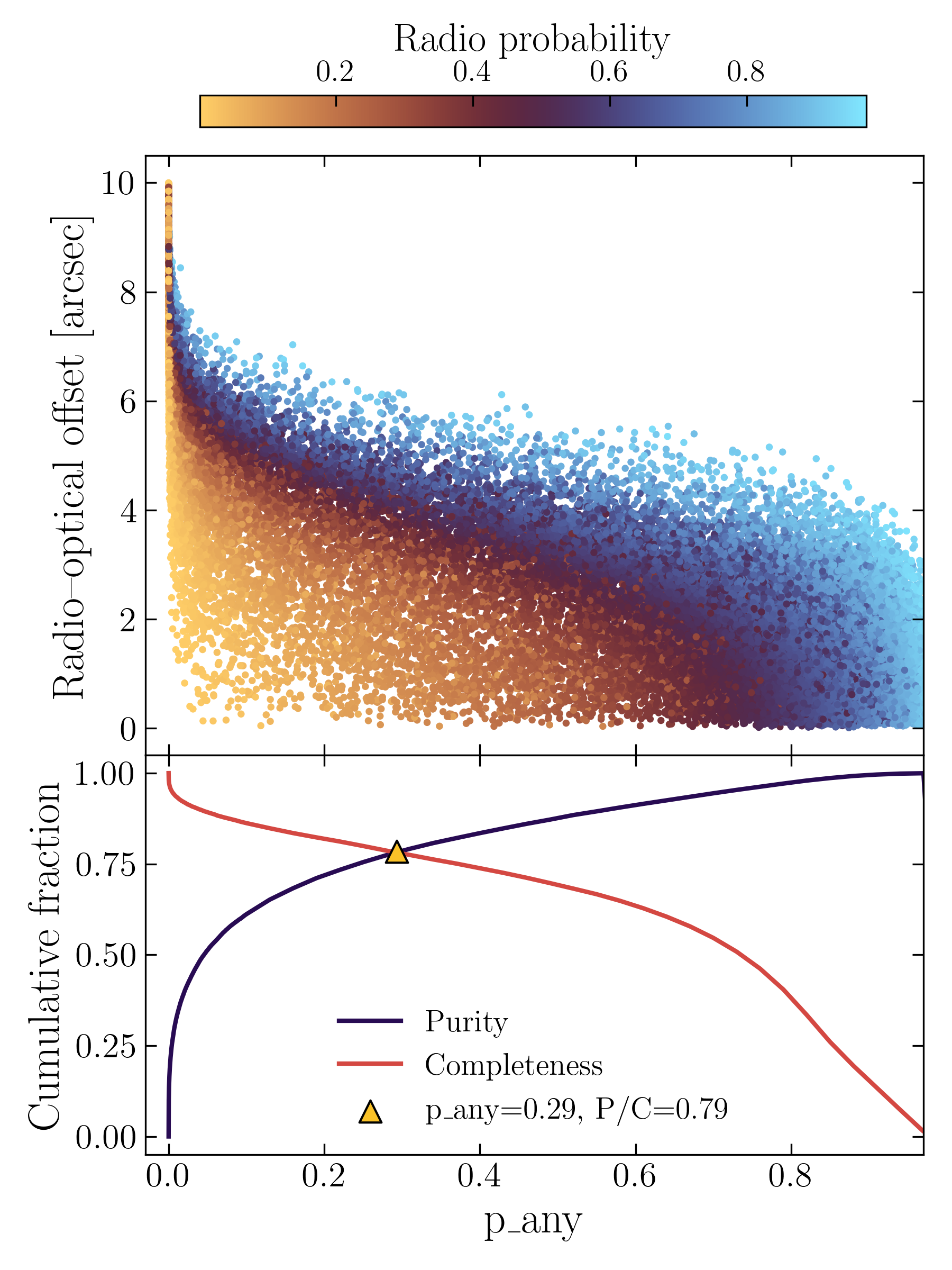}
\caption{Top: Radio–optical separation as a function of association probability (p\_any) for all candidate counterparts, colour-coded by the probability of being a radio emitter. Bottom: Cumulative purity and completeness curves as a function of p\_any. The intersection point at $\textrm{p\_any} \approx 0.29$ (black triangle) marks the optimal balance between purity and completeness ($\sim$80\%).}
\label{fig:3}
\end{figure}

Other reasons why an optical CTP could not be identified include: 
(1) proximity to the Small or Large Magellanic Clouds, where LS10 entries are limited to brighter sources identified by Gaia \citep[][]{Gaia2016}; (2) incomplete coverage in the redder LS10 bands; (3) edge effects near the LS10 footprint boundaries; (4) regions affected by bright star saturation and other imaging artefacts and (5) significant radio–optical positional offsets. The latter cannot be reliably addressed by simply enlarging the matching radius, as this would increase the risk of false associations. In fact, when restricting the search radius in {\sc NWAY} to 3 arcseconds, only $\sim$ 33\,500 CTPs remain, with fewer than 5\% of these switching their primary match, presumably because the true CTP lies outside this radius. 

To robustly estimate the fraction of radio galaxies at $z \gtrsim 1$, it is essential to quantify the rate of chance associations and assess the reliability of low-probability CTP candidates within the adopted search radius. To this end, we repeated the matching procedure for a control sample of FLASH continuum sources whose sky coordinates were artificially shifted, thereby representing random associations. The resulting cumulative distributions of the CTP identification probability (p\_any) for both the real and shifted samples yield the purity–completeness relation as a function of p\_any, shown in the lower panel of \Cref{fig:3}. By maximizing both quantities simultaneously, we find that adopting a lower threshold of p\_any $\sim 0.3$ achieves roughly 80\% purity and completeness, corresponding to an estimated chance-association rate of $\sim 20\%$. As illustrated in the upper panel of \Cref{fig:3}, this threshold also naturally limits the radio–optical positional offsets to to a maximum of 6 arcseconds for sources with a high probability of being a radio emitter according to the RACS-based prior. Consequently, throughout the remainder of this work we restrict our analysis to CTP associations with
p\_any $ \ge 0.292$  that are detected in all LS10 optical bands, yielding a final sample of approximately 27\,500 CTPs.

\section{Radio AGN diagnostics}
\label{app:2}

Two commonly used approaches to identify AGN in the radio regime are based on RL and radio-excess (REX), where each method probes emission ratios as a tracer of different underlying physical properties \citep[][]{Yun01, Delmoro13, Drake24, Mazzo_2026}. The RL parameter is traditionally defined as 
\begin{equation}
    \rm{RL} = \frac{\it{f}_{\rm{radio}}}{\it{f}_{\rm{optical}}}
\end{equation}
and is designed to identify AGN whose emission is dominated by relativistic jets in the radio compared to the optical produced by the accretion disk or host galaxy \citep[][]{Kellermann_2016}. This diagnostic is therefore most sensitive to separating AGN into RL and RQ classes within an already AGN-dominated population. However, it is less effective in discriminating AGN from SFGs, since the observed optical emission must not necessarily be linked to the processes responsible for SF-related radio emission. By contrast, the REX approach attempts to disentangle AGN from SFGs, by comparing the observed radio luminosity with the expected contribution from SF processes \citep[][]{Donley_2005,Panessa_2019,Igo_2024}. This usually requires a reliable estimate of the star formation rate (SFR) and consequently the associated radio emission. In practice, this is often achieved through SED template fitting, where the host galaxy’s SF radio component is modelled and subtracted to estimate an AGN radio excess. However, template fitting carries its own uncertainties, particularly if the underlying SF history, dust attenuation, or non-thermal emission processes are poorly constrained. Alternatively, following the definition of

\begin{equation}
    \log L_{144\, \rm{MHz}} = 14 - \frac{M_{W3}}{2.5}
\end{equation}
from \cite{hardcastle2025} to separate AGN and SFGs, we retrieve an AGN fraction that exceeds 99.5\%, suggesting that contamination from SFGs in our sample is negligible. Another widely used empirical proxy for estimating the expected radio luminosity from SF is the far-IR (FIR)–radio correlation \citep[FIRC,][]{Helou_1985,Condon_1992,Delhaize_2017}. This relation connects the FIR emission from dust heated by young stars with the synchrotron emission from supernova-driven cosmic rays, both of which trace recent SF. This method has the advantage of physically removing SFG contaminants from an otherwise mixed population, though it depends critically on the availability and validity of the FIR–radio relation at the redshift and luminosity regime in question \citep[][]{Wang_2024}. A more simplified method for identifying radio AGN is therefore to select sources whose radio emission exceeds the level expected from SF alone. Recent deep studies \citep[e.g.][]{Wang_2024} place the cross-over luminosity $L$, i.e. the threshold beyond which the number density of radio AGN surpasses that of SFGs, at $L_{1.4\,\rm{GHz}}\gtrsim10^{23}$ W Hz$^{-1}$ at $z \sim 0$, rising with redshift up to $L_{1.4\,\rm{GHz}}\gtrsim10^{25}$ W Hz$^{-1}$ by $z \sim 4$ (refer to \Cref{fig:9}). Although many radio sources appear to have a power-law dominated SED across a broad range of radio frequencies, the slope of this power-law may be either `steep' or `flat' and some sources show more complex behaviour with spectral peaks or troughs \citep[see e.g.][]{dezotti2010,Kerrison24}. As a result, measurements made at one radio frequency cannot typically be reliably extrapolated to frequencies that are significantly higher or lower. However, to quantify the intrinsic radio power of our sources, we calculate their radio luminosities following the standard convention:

\begin{equation}
    L_{856\rm{MHz}}\,[{\rm{W}\, \rm{Hz}^{-1}}] = 4 {\pi} \, d^{2}_{L} \, S_{\nu} \, 10^{-30} \, (1+z)^{\alpha -1} \, ,
\end{equation}
where $d_{L}$ is the luminosity distance in cm, $S_{\nu}$ is the total integrated flux in units of Jansky (Jy) and $(1+z)^{\alpha -1}$ is the K-correction, assuming a radio spectral index $\alpha = 0.7$ \citep[][]{Condon_1992}. Given $\alpha$, we can convert $L_{856\rm{MHz}}$ into the more conventionally used $L_{1.4\rm{GHz}}$ by

\begin{equation}
    L_{1.4\rm{GHz}} =  L_{856\rm{MHz}} \bigg (\frac{1.4}{0.856}\bigg )^{-\alpha} \, .
\end{equation}

\section{Catalogue column description}
\label{app:3}

\begin{enumerate}
 \item FLASH\_COMPONENT\_ID: Unique source ID assigned to each FLASH source
  \item FLASH\_SBID: FLASH field ID
  \item Columns 3--4: ASKAP-FLASH coordinates and respective positional uncertainty
  \item Columns 5--7: ASKAP-FLASH peak and integrated fluxes
  \item Columns 8--10: Binary flag indicating the (number of) repeated entries of a source and the maximum group separation due to observations of overlapping fields
  \item LS10\_FULLID: Unique LS10 source ID assigned to optical counterpart. It is created by concatenating the LS10 coloumns RELEASE, BRICKID and OBJID.
  \item Columns 12--13: Right Ascension and Declination in degrees of the LS10 optical counterpart
  \item Columns 14--15: Binary flags indicating the observation of an LS10 source in all or any of the optical $g,r,i,z$ bands
  \item LS10\_type: LS10 morphological classification\footnote{\url{https://www.legacysurvey.org/dr10/catalogs/}}
  \item Columns 17--24: LS10 fluxes divided by Milky Way transmission
  \item LS10\_snr: Binary flag indicating whether the signal to noise ratio is < 3 for any of the optical $g,r,i,z$ 
  \item Columns 26--34: NWAY output as presented in the Appendix of \cite{Salvato2018}
  \item Columns 35--36: LS10 counterpart sample purity and completeness evaluated at p\_any per source
  \item PICZL\_z\_phot: {\sc{PICZL}} photo-$z$ estimate (most prominent PDF mode)
  \item PICZL\_pdf\_degeneracy: Classification of photo-$z$ PDF degeneracy based on the presence, prominence, and separation of (one or more) secondary peaks relative to the primary PDF peak. Values range from `none' (single-peaked PDFs) through `light' and `medium', to `strong' degeneracy
  \item \label{itm:sec_peak_psf} PICZL\_secondary\_peak\_PSF: Secondary peak at $z_{\rm{phot}} < 1.5$ for sources of non-PSF type where $z_{\rm{phot}} > 1.5$ when available
  \item \label{itm:sec_peak_high_z} PICZL\_secondary\_peak\_high\_z: Secondary peak at $z_{\rm{phot}} < 2$ for sources of any type where $z_{\rm{phot}} > 2$ when available
  \item PICZL\_best\_z\_phot: \Cref{itm:sec_peak_psf} when available for sources with $z_{\rm phot} > 1.5$, otherwise \Cref{itm:sec_peak_high_z} for sources with $z_{\rm phot} > 4.7$ when available
  \item Columns 42--43: upper and lower $1 \sigma$ errors, defined as the highest posterior density (HPD) interval
  \item PICZL\_multi\_peak\_errors: Binary flag indicating whether the $1 \sigma$ error HPD interval spans multiple peaks
\end{enumerate}

\end{document}